\documentclass[12pt]{article}

\RequirePackage{amsthm,amsmath,amsfonts,amssymb}
\RequirePackage[authoryear]{natbib}
\RequirePackage{graphicx}
\RequirePackage{booktabs}
\RequirePackage{enumitem}
\RequirePackage{xcolor}
\RequirePackage[margin=1in]{geometry}
\RequirePackage{setspace}
\graphicspath{{figures/}}

\usepackage[hidelinks]{hyperref}
\usepackage{xurl}

\newtheorem{proposition}{Proposition}

\begin{document}

\title{HaploPerturb: Low-rank copula construction of haplotype perturbations improves sequence-to-function analysis of Alzheimer's disease loci}
\author{Jichun Xie\\
\normalsize Department of Biostatistics and Bioinformatics, Duke University\\
\normalsize Department of Mathematics, Duke University\\
\normalsize \texttt{jichun.xie@duke.edu}}
\date{}
\maketitle

\begin{abstract}
Sequence-to-function models predict molecular phenotypes from complete sequence windows. At
genome-wide association study loci, however, the prevailing design perturbs only the lead variant
on the reference genome, even though the lead is often correlated with nearby variants through
linkage disequilibrium. This single-variant perturbation implicitly fixes all linked alleles at their
reference-genome states and may therefore create an uncommon or unobserved population haplotype. We
study this input-construction problem at 38 Alzheimer's disease loci. We introduce HaploPerturb,
which uses phased ROS/MAP genotypes or the European 1000 Genomes panel to fit a fixed-margin latent
Gaussian factor model and rank partner configurations conditional on each lead allele. The leading
public-panel construction agrees with the donor-panel construction at all loci under a strict
linkage-disequilibrium threshold and 36 of 37 loci under a broader threshold after restricting to
shared partners. Known-truth simulations show exact recovery of the dominant configuration under
strong linkage disequilibrium and expose persistent residual correlation under a misspecified
one-factor model. In an AlphaGenome benchmark against cell-type-specific ROS/MAP eQTLs, broader-set
public-panel haplotypes yield microglial enrichment of 2.07 (95\% whole-locus bootstrap percentile
interval 1.48--3.60), compared with 1.43 (0.69--2.29) for a lead-only edit. Empirical-mode and
LD-sign backgrounds yield 2.20 (1.60--3.64), with no detectable advantage or loss relative to the
HaploPerturb top configuration. Thus population-informed sequence construction matters in this
application, while the choice among reasonable leading haplotype rules is less consequential than
the choice between a haplotype and a lead-only reference background.
\end{abstract}

\medskip
\noindent\textbf{Keywords:} linkage disequilibrium; Gaussian copula; factor model;
sequence-to-function model; expression quantitative trait locus.

\section{Introduction}
\label{sec:intro}

Sequence-to-function models predict molecular phenotypes from a complete DNA sequence. At a
genome-wide association study (GWAS) locus, however, the usual input changes only the lead variant
on the reference genome. Every linked allele is silently left at its reference state. The resulting
sequence is easy to generate, but it need not represent a haplotype carried by the population being
studied. This is not a detail internal to the prediction model: Enformer, Borzoi and AlphaGenome read
long genomic windows, so changing the surrounding allelic background can change the predicted
effect even when the model, output track and lead allele are held fixed
\citep{avsec2021effective,linder2025predicting,avsec2026advancing}.

The problem is concrete in Alzheimer's disease. Its GWAS loci contain long and sometimes dense
linkage-disequilibrium (LD) blocks, and functional interpretation increasingly uses cell-type
specific sequence predictions. Existing benchmarks compare models, tasks and evaluation sets
\citep{huang2023personal,sasse2023benchmarking,feng2025benchmarking,de2025annotating}, but usually
take the supplied perturbation as given. We ask a prior question: after fixing either allele of an
Alzheimer's disease lead variant, which partner-allele configuration should be written into the
prediction window? A useful answer must be population-informed, computationally feasible at loci
with thousands of partners, and explicit about uncertainty when a finite panel cannot resolve exact
high-dimensional haplotypes.

We analyze 38 lead variants from the Wightman et al. GWAS \citep{wightman2021genome}, using phased
genotypes from ROS/MAP and the European 1000 Genomes panel. Our framework, HaploPerturb, models
high-LD partners with a fixed-margin latent Gaussian factor model. Conditioning on either lead
allele yields a distribution over partner configurations; an exact factor-space sampler discovers
candidates, which are then ranked by HaploPerturb probability. We compare the HaploPerturb
construction with three prespecified inputs:
the observed conditional mode, a background based only on the sign of lead--partner LD, and the
usual single-lead edit on GRCh38. AlphaGenome RNA predictions are benchmarked against cell-type
specific ROS/MAP \emph{cis}-eQTLs from paired whole-genome and single-nucleus RNA sequencing
\citep{fujita2024cell}.

Three applied findings organize the paper. First, after high-LD selection, one factor captures most
latent variance, and leading HaploPerturb configurations are observed frequently even though lower-ranked
exact haplotypes are sparsely repeated in the finite panels. Controlled simulations confirm that
finite panel size, rather than evidence of zero population probability, explains many absent lower
configurations. Second, the leading public- and donor-panel constructions agree at all loci under
the strict partner threshold and at 36 of 37 loci under the broader threshold after comparison on
shared partners. Third, in the broader arm, population-informed public-panel backgrounds improve
microglial eQTL ranking relative to a lead-only edit. The HaploPerturb, empirical-mode and LD-sign
backgrounds perform similarly; the evidence supports haplotype-aware input construction, not
superiority of the factor model over every simple top-one rule. Astrocyte and excitatory-neuron
analyses show no corresponding enrichment, a result that may reflect prediction accuracy,
biosample correspondence, eQTL power or the RNA summary rather than any single limitation of
AlphaGenome.

Statistically, HaploPerturb supplies more than a single heuristic background: it defines the
conditional estimand, supports several ranked or averaged inputs, and provides diagnostics and
candidate-coverage checks for difficult loci. Scientifically, holding AlphaGenome fixed isolates a
previously hidden analysis choice and shows that it changes prioritization across established
Alzheimer's disease loci. Section~\ref{sec:data} introduces the data and partner arms;
Sections~\ref{sec:construction}--\ref{sec:estimation} define and estimate the conditional model;
Section~\ref{sec:haplotype-analysis} validates the construction empirically and by simulation;
Section~\ref{sec:application} reports the Alzheimer's disease eQTL application; and
Section~\ref{sec:discussion} gives implementation guidance, limitations and extensions.

\section{Data overview}
\label{sec:data}

Our pipeline starts from the 38 Alzheimer's disease GWAS lead variants reported by \citet{wightman2021genome}.
For each lead, the 1000 Genomes European panel defines its phased LD partners; ROS/MAP supplies cohort genotypes and the eQTL benchmark; and AlphaGenome predicts the regulatory effects of the resulting sequence contrasts.
Throughout, each locus identifier contains four colon-separated fields: chromosome, 1-based GRCh38 position, GRCh38 reference allele and alternate allele. For example, \texttt{17:46062125:A:C} denotes an A-to-C variant at position 46,062,125 on chromosome 17.

\paragraph*{Reference haplotype panel}
The 1000 Genomes Project 30$\times$ high-coverage GRCh38 release contains $3{,}202$ individuals: the $2{,}504$ mutually unrelated individuals from Phase 3 and $698$ additional relatives \citep{byrska2022high, clarke20121000}. Its European superpopulation contains $633$ individuals from CEU, FIN, GBR, IBS and TSI. The nested unrelated European panel contains $503$ individuals.

The legacy lead--partner screen used all $633$ individuals. For each GWAS lead, candidate variants were taken from an approximately $\pm524$-kb window, after requiring the stored 1000 Genomes annotations $\mathrm{MAF}_{\mathrm{EUR,unrel}}\geq0.01$ and $\mathrm{HWE}_{\mathrm{EUR}}\geq10^{-6}$. PLINK~2 \texttt{--r-phased} then used the two observed phased alleles from each individual, for $1{,}266$ haplotypes in total. If $X_{hj}\in\{0,1\}$ is PLINK's allele indicator for variant $j$ on haplotype $h$, the lead--candidate statistic was the squared sample Pearson correlation
      \[
      \widehat r_{0j}^{\,2}
      =
      \left[
      \frac{\sum_{h=1}^{1266}(X_{h0}-\overline X_0)(X_{hj}-\overline X_j)}
      {\{\sum_{h=1}^{1266}(X_{h0}-\overline X_0)^2
          \sum_{h=1}^{1266}(X_{hj}-\overline X_j)^2\}^{1/2}}
      \right]^2 .
      \]
Reversing the allele coding at either variant changes the sign of $\widehat r_{0j}$ but not $\widehat r_{0j}^{\,2}$, so partner selection is intentionally sign-blind. The query stage stored pairs with $\widehat r_{0j}^{\,2}\geq0.1$. The resulting locus--partner manifests were kept fixed. All alternate-allele frequencies, signed phased correlations, haplotypes and factor-model fits reported downstream were recomputed on the nested $503$ unrelated individuals, giving $1{,}006$ phased haplotypes.

\begin{figure}[t]
\centering
\includegraphics[width=0.83\textwidth]{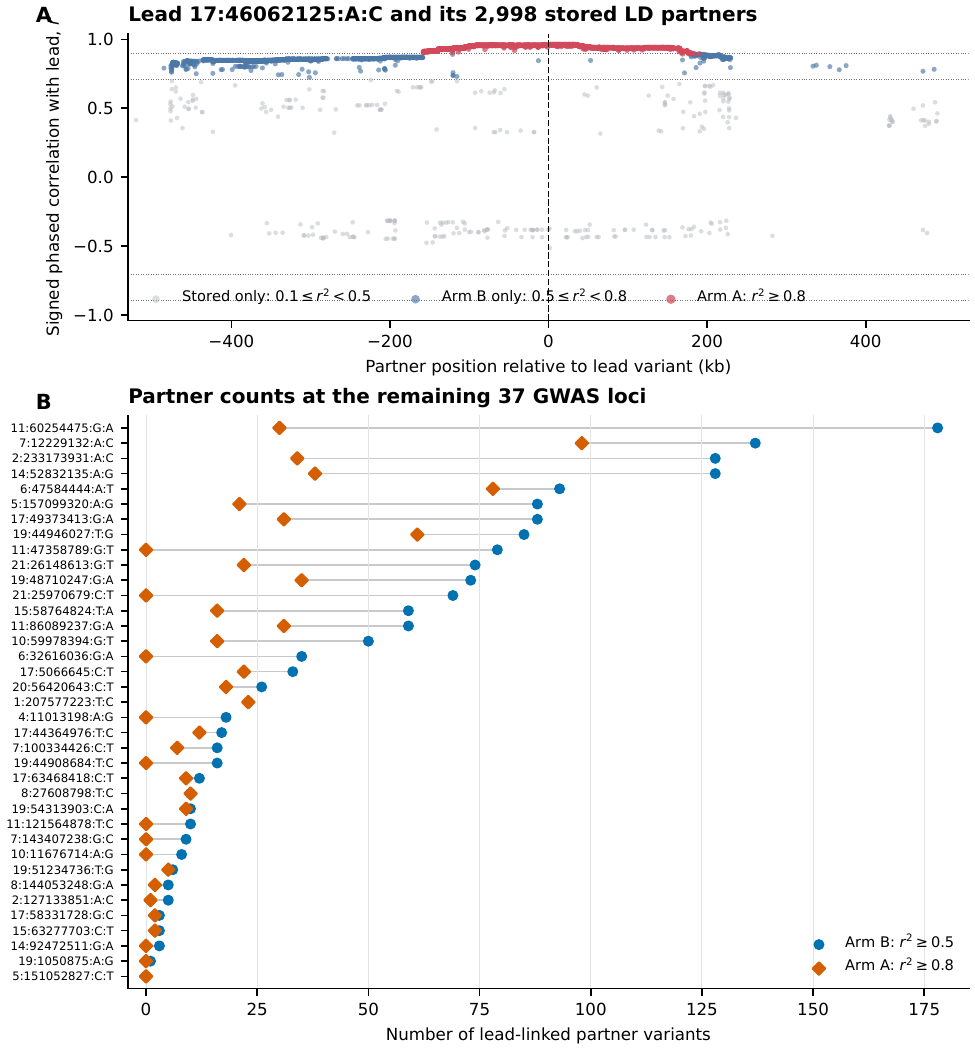}
\caption{\textbf{Local phased LD and the number of partner variants retained at each GWAS locus.} Arm A retains partners with $r^2\geq0.8$, whereas Arm B uses the broader threshold $r^2\geq0.5$. \textbf{(A)} Local LD around lead variant \texttt{17:46062125:A:C} in the 1000 Genomes EUR panel. Each point is a candidate partner; horizontal position is distance from the lead and vertical position is the signed phased correlation between alternate-allele indicators. Gray points fall below the Arm B threshold ($0.1\leq r^2<0.5$), blue points enter Arm B only ($0.5\leq r^2<0.8$), and red points enter both arms ($r^2\geq0.8$). The vertical dashed line marks the lead, and the horizontal dotted lines mark the signed correlation values corresponding to the two thresholds. This locus contributes $2{,}693$ partners to Arm B and $1{,}285$ to Arm A. \textbf{(B)} Partner counts for the other $37$ GWAS loci, ordered from largest to smallest Arm B count. Blue circles and orange diamonds show the Arm B and Arm A counts, respectively; a gray line connects the two values for the same locus. The chromosome 17 locus shown in (A) is omitted from (B) to keep the remaining counts legible.}
\label{fig:ld-example}
\end{figure}

\paragraph*{GWAS loci and partner arms}
The 38 GWAS leads define 38 sequence windows. All were observed and variable in the 503-person 1000 Genomes European panel. We compare two partner thresholds to assess how the size of the retained LD structure affects haplotype construction; signed $\widehat r_{0j}$ determines allelic alignment after the partner set is selected by $\widehat r_{0j}^{\,2}$.
      \begin{itemize}
\item \emph{Arm A}, $r^2 \ge 0.8$: $27$ windows carry at least one partner, $1{,}918$ partners in all, and \texttt{17:46062125:A:C} is the lead variant with the largest partner set ($k = 1{,}285$ partners). This window is the only one with $p=k+1$ exceeding the $n = 1{,}006$ haplotypes it is estimated from.
\item \emph{Arm B}, $r^2 \ge 0.5$: $37$ windows carry at least one partner, $4{,}350$ partners in all, and \texttt{17:46062125:A:C} is the lead variant with the largest partner set ($k = 2{,}693$ partners).
      \end{itemize}
Figure~\ref{fig:ld-example} shows the largest partner set and the counts at the remaining loci. Loci without a partner at a given threshold reduce to a single-lead-variant flip and do not require the construction in Section~\ref{sec:construction}.

\paragraph*{ROS/MAP genotypes and benchmark eQTLs}

ROS/MAP serves two roles: donor genotypes provide the primary cohort-specific factor-model fit \citep{bennett2018religious}, and cell-type \emph{cis}-eQTLs provide the benchmark. After comparing the ROS/MAP manifest with 1000 Genomes, we found that not every selected partner was available in ROS/MAP. Arm A misses $45$ partners across $5/27$ loci; Arm B misses $152$ partners across $12/37$ loci. The ROS/MAP fits omitted unavailable partners; none was imputed or replaced using a 1000 Genomes haplotype.

For eQTL mapping, \citet{fujita2024cell} paired whole-genome-sequencing genotypes with dorsolateral-prefrontal-cortex single-nucleus RNA-seq profiles from the same 424 ROS/MAP participants, totaling approximately 1.5 million nuclei. We obtained the full summary statistics through AD Knowledge Portal/Synapse record \texttt{syn52335732} and use the astrocyte, excitatory-neuron and microglial files. We focus on these three cell types because the predefined panel of 40 AlphaGenome biosamples contains corresponding tracks or usable surrogates for them; the other Fujita cell types have no appropriate proxy in that panel.

Within each cell type, Fujita et al. tested common variants within 1 Mb of a gene's transcription start site. Their two-step correction first Bonferroni-adjusted each gene's smallest nominal SNP $P$ value for the number of cis-SNPs tested for that gene, then applied the Benjamini--Hochberg procedure to those gene-wise values at a 5\% FDR. They set the eSNP nominal-$P$ threshold to the largest nominal $P$ value among the resulting significant gene--SNP pairs; the released \texttt{significant\_by\_2step\_FDR} column records this call. This is a useful benchmark because it supplies published, multiplicity-controlled cell-type-specific associations based on paired genotypes and measured expression, defined independently of AlphaGenome and our ranking scores. A locus--gene pair is benchmark-positive only if a variant--gene pair within the same AlphaGenome input window carries this indicator. Because the factor-model fit and benchmark both use ROS/MAP, the comparison is within cohort rather than an independent replication.

\section{HaploPerturb construction by low-rank factorization}
\label{sec:construction}

\subsection{The perturbation problem}
\label{sec:problem}

Let $X = (X_0,\dots,X_k) \in \{0,1\}^{k+1}$ denote the perturbed haplotype, where $X_0$ is the lead variant, and $X_j$ with $j\in[k]$ are partner variants. For any $j\in\{0\}\cup [k]$, let $X_j=1$ code the alternate allele and $X_j=0$ code the reference allele with respect to the GRCh38 primary assembly.

Querying a sequence-to-function model requires three choices: which partners enter the perturbation, which allele each carries, and what background sequence surrounds them. Section~\ref{sec:data} specifies the partner sets. We use the GRCh38 primary assembly as the surrounding background to facilitate practical use, with its implications and alternatives discussed in Section~\ref{sec:discussion}. Selecting the partner alleles is the statistical problem studied in this section.

Define the conditional law
\begin{equation}
\Pr\bigl(X_1 = x_1, \dots, X_k = x_k \;\bigm|\; X_0 = s\bigr),
\qquad s \in \{0,1\}.
\label{eq:target}
\end{equation}
Our goal is to find the haplotypes whose conditional probability ranks at the top of it. Both values of $s$ are of interest because an analyst may wish to perturb the lead to either allele.

Hereafter, HaploPerturb refers to the fitted conditional distribution together with the procedure
used to discover candidates, evaluate their probabilities and report their ordering. A
HaploPerturb rank is a configuration's position in that ordering; HaploPerturb top one is the
leading configuration, and HaploPerturb top $L$ uses the first $L$ configurations.

\subsection{A latent Gaussian model with low-rank factors}
\label{sec:model}

We model $X$ through a latent Gaussian vector, the standard Gaussian-copula device for correlated binary data \citep{nelsen2006copulas, xue2000multivariate, inouye2017review}. Let $p_j \in (0,1)$ be the alternate-allele frequency of variant $j$ in the reference population represented by the panel and set
\begin{equation}
Z = (Z_0,\dots,Z_k)^{\!\top} \sim \mathcal{N}_{k+1}(0, \mathbf{R}),
\qquad
X_j = \mathbf{1}\{Z_j > \tau_j\},
\qquad
\tau_j = \Phi^{-1}(1-p_j),
\label{eq:latent}
\end{equation}
where $\mathbf{R}$ is a correlation matrix, $\Phi$ and $\phi$ are the standard normal distribution and density functions, and $\mathbf{1}\{\cdot\}$ is the indicator. Collect the thresholds into $\tau = (\tau_0,\dots,\tau_k)^{\!\top}$, so that \eqref{eq:latent} reads $X = \mathbf{1}\{Z > \tau\}$ coordinatewise. Then $\Pr(X_j = 1) = p_j$ exactly, so the model reproduces the reference-population allele frequencies without further calibration. Write $p = k+1$ for the number of variants, and index the panel's phased haplotypes by $i = 1,\dots,n$, so $X_{ij}$ is the allele of variant $j$ on haplotype $i$; a panel of $n/2$ individuals contributes $n$ haplotypes.

Enumerating \eqref{eq:target} directly is infeasible --- the support has $2^k$ points and $k$ could reach hundreds, thousands, or more here. Fortunately, our empirical analysis showed that $\mathbf{R}$ usually has a low-rank factor structure:
\begin{equation}
Z_{ij} \;=\; \mathbf{b}_j^{\!\top} f_i + \varepsilon_{ij},
\qquad
f_i \sim \mathcal{N}_q(0, I_q),
\qquad
\varepsilon_i \sim \mathcal{N}_{p}(0,\mathbf{D}),
\qquad
\mathbf{D} = \operatorname{diag}(\psi_0,\dots,\psi_k),
\label{eq:latent-factor}
\end{equation}
with $f_i$ the common factor, $\mathbf{b}_j \in \mathbb{R}^{q}$ the $j$th row of a loading matrix $\mathbf{B} \in \mathbb{R}^{p\times q}$, $\varepsilon_i$ independent of $f_i$, and $\psi = (\psi_0,\dots,\psi_k)^{\!\top}$ the vector of residual variances, so $\mathbf{D} = \operatorname{diag}(\psi)$. Requiring unit variance at every variant,
\begin{equation}
\lVert \mathbf{b}_j\rVert^2 + \psi_j = 1
\qquad\text{for every } j = 0,1,\dots,k,
\label{eq:unitvar}
\end{equation}
which is $p$ separate scalar constraints, keeps $Z$ a correlation and gives
\begin{equation}
\mathbf{R} \;=\; \mathbf{B}\mathbf{B}^{\!\top} + \mathbf{D}.
\label{eq:factor}
\end{equation}

Two properties matter later:
\begin{itemize}
\item $\mathbf{B}\mathbf{B}^{\!\top}+\mathbf{D}$ is positive semi-definite at every parameter value; and
\item $\mathbf{B}$ is identified only up to an orthogonal transformation. Write $O(q) = \{Q \in \mathbb{R}^{q\times q} : Q^{\!\top}Q = I_q\}$. If $f \sim \mathcal{N}_q(0,I_q)$ then $Qf \sim \mathcal{N}_q(0, QQ^{\!\top}) = \mathcal{N}_q(0,I_q)$, so replacing $(\mathbf{B}, f)$ by $(\mathbf{B}Q, Q^{\!\top}f)$ leaves $\mathbf{B}f$ and hence the law of $X$ unchanged. This costs nothing: every quantity we report --- $\mathbf{B}\mathbf{B}^{\!\top}$, $\psi$, the thresholds, haplotype probabilities and their ranking --- is invariant under $O(q)$. Note that $O(q)$ includes reflections as well as proper rotations, and it is the reflections that matter below.
\end{itemize}

We adopt \eqref{eq:latent-factor} because, at any $q$, the variants are independent conditionally on the factor,
\begin{equation}
q_j(f) \;:=\; \Pr\bigl(X_{ij} = 1 \mid f_i = f\bigr)
\;=\; \Phi\!\left(\frac{\mathbf{b}_j^{\!\top} f - \tau_j}{\sqrt{\psi_j}}\right).
\label{eq:qj}
\end{equation}
Equation \eqref{eq:qj} runs from the factor to the alleles; the construction needs it in both directions. Write the lead's conditional likelihood at either allele as
\[
\ell_0(f;s) \;:=\; \Pr\bigl(X_{i0} = s \mid f_i = f\bigr)
\;=\; q_0(f)^{s}\bigl\{1-q_0(f)\bigr\}^{1-s},
\qquad s \in \{0,1\}.
\]
Combining this with the prior $f_i \sim \mathcal{N}_q(0,I_q)$, conditioning on the lead reweights the factor by Bayes' rule,
\begin{equation}
p\bigl(f \mid X_{i0} = s\bigr) \;=\; \frac{\phi_q(f)\,\ell_0(f;s)}{\pi_0(s)},
\qquad
\pi_0(s) \;=\; \Pr(X_{i0} = s) \;=\; p_0^{\,s}(1-p_0)^{1-s},
\label{eq:factorpost}
\end{equation}

Given the factor the partners are still independent, by \eqref{eq:qj}, so for either allele of the lead
\begin{equation}
\Pr\bigl(X = x \mid X_0 = s\bigr)
\;=\; \int_{\mathbb{R}^q} p\bigl(f \mid X_0 = s\bigr)
\prod_{j=1}^{k} q_j(f)^{x_j}\bigl\{1-q_j(f)\bigr\}^{1-x_j}\,\mathrm{d}f ,
\qquad s \in \{0,1\}.
\label{eq:condlaw}
\end{equation}
The $2^k$ sum in \eqref{eq:target} has become a $q$-dimensional integral, and Section~\ref{sec:closedform} describes how we search its high-probability configurations. The lead enters \eqref{eq:condlaw} only through the weight $p(f \mid X_0 = s)$: the partner factors are the same functions of $f$ at both values of $s$. This is why the two conditionings give genuinely different laws over the partners rather than complements of one another, and why both must be constructed separately when both perturbations are of interest.

At $q = 1$ each $\mathbf b_j$ is a scalar, so its sign is meaningful --- at $q > 1$ it is a vector and no sign exists. The following proposition says the loading vector is then determined up to that sign alone, fixes it by orienting the factor on the lead, and identifies what the sign means.

\begin{proposition}[Identifiability at one factor]
\label{prop:ident}
Take $q = 1$ in \eqref{eq:latent-factor} and let the admissible parameter space be
\[
\Theta \;=\; \bigl\{(\mathbf b,\tau) \in \mathbb{R}^{p}\times\mathbb{R}^{p} :\;
\psi_j = 1-b_j^2 \in (0,1] \ \text{for every } j \bigr\},
\]
so that all thresholds are finite and every marginal satisfies $0 < \Pr(X_j = 1) < 1$. Let $\mathcal{L}(\mathbf b,\tau)$ denote the induced law of $X$. For $(\mathbf b,\tau)$ and $(\mathbf b',\tau')$ both in $\Theta$:
\begin{enumerate}[leftmargin=*,itemsep=1pt]
\item[(i)] $\mathcal{L}(-\mathbf b,\tau) = \mathcal{L}(\mathbf b,\tau)$;
\item[(ii)] if at least three coordinates of $\mathbf b$ are non-zero and $\mathcal{L}(\mathbf b',\tau') = \mathcal{L}(\mathbf b,\tau)$, then $\tau' = \tau$ and $\mathbf b' = \pm\mathbf b$;
\item[(iii)] if in addition $b_0 \ne 0$, the normalization $b_0 > 0$ selects a unique representative, and for every \emph{partner} $j \ne 0$ with $b_j \ne 0$, $\operatorname{sign}(b_j) = \operatorname{sign}(R_{0j})$.
\end{enumerate}
\end{proposition}

Part (i) says the choice of sign is free. Part (ii) says nothing further is left to choose. Part (iii) is why the normalization is worth stating: it makes the loading sign \emph{equal} to $\operatorname{sign}(R_{0j})$, the sign of the latent correlation between partner $j$ and the lead. A positive value means variant $j$'s alternate allele travels with the lead's alternate allele, a negative value that it travels with the lead's reference allele. Supplementary Note~3 gives the proof, using the strict monotonicity of the bivariate-normal probability in its correlation parameter \citep{Plackett1954}.

\subsubsection{A working parameterization.} Constraint \eqref{eq:unitvar} makes $(\mathbf B,\tau)$
awkward to optimize directly; dividing the probit argument by $\sqrt{\psi_j}$ absorbs it exactly. Put
\begin{equation}
a_j = \frac{\mathbf b_j}{\sqrt{\psi_j}} \in \mathbb{R}^{q},
\qquad
\beta_j = -\frac{\tau_j}{\sqrt{\psi_j}} \in \mathbb{R},
\qquad\text{so}\qquad
q_j(f) = \Phi\bigl(a_j^{\!\top} f + \beta_j\bigr).
\label{eq:ogive}
\end{equation}
The map back is closed form: $\psi_j = (1+\lVert a_j\rVert^2)^{-1}$, $\mathbf b_j = a_j\sqrt{\psi_j}$, $\tau_j = -\beta_j\sqrt{\psi_j}$, and $\lVert\mathbf b_j\rVert^2 + \psi_j = 1$ holds identically. The likelihood contributed by one observed haplotype $x_i \in \{0,1\}^{p}$ is then
\begin{equation}
L\bigl(x_i; a,\beta\bigr) \;=\;
\int_{\mathbb{R}^q} \phi_q(f) \prod_{j=0}^{k}
q_j(f)^{x_{ij}}\bigl\{1-q_j(f)\bigr\}^{1-x_{ij}}\,\mathrm{d}f ,
\label{eq:marglik-main}
\end{equation}
with $\phi_q$ the $\mathcal{N}_q(0,I_q)$ density. The reparameterization is what makes \eqref{eq:marglik-main} tractable to maximize: \eqref{eq:unitvar} is absorbed rather than imposed, so no repair step is needed to keep the latent variances at one.

The parameterization is not, however, unconstrained in practice. Nothing above prevents $\lVert a_j\rVert \to \infty$, which is $\psi_j \to 0$: variant $j$ becomes a deterministic function of the factor, and the likelihood can approach its supremum on that boundary. Fitting therefore imposes a floor $\psi_j \ge \psi_{\min} > 0$, which by the map back is the norm-ball constraint $\lVert a_j\rVert^2 \le \psi_{\min}^{-1} - 1$.

\subsection{Finding high-probability haplotypes}
\label{sec:closedform}

Ranking \eqref{eq:condlaw} directly would require $2^k$ probability integrals. At $k=20$, one list already requires more than a million evaluations, and our largest locus has thousands of partners. We therefore consider two strategies for constructing a smaller set of plausible high-probability haplotypes before evaluating their HaploPerturb probabilities: model-guided conditional sampling and deterministic enumeration of factor-space modes. The two strategies use the same HaploPerturb conditional law but make different compromises. Sampling searches broadly and scales to large loci and multiple factors, at the cost of random candidate discovery and probabilistic rather than deterministic coverage. Factor-space enumeration is reproducible and computationally efficient when the number of regions is small, but its candidate set is structurally restricted and becomes impractical as the factor dimension increases. Neither strategy by itself guarantees the global top-$L$ haplotypes.

\subsubsection{Model-guided conditional sampling}
\label{sec:conditional-sampling}

The sampling algorithm is easiest to see at $q=1$. Unit latent variance gives $Z_0\sim\mathcal N(0,1)$. To condition on a chosen lead state $s$, draw $Z_0$ from this normal distribution truncated to $(\tau_0,\infty)$ when $s=1$ or to $(-\infty,\tau_0]$ when $s=0$. Given $Z_0=z_0$, draw the scalar factor from $f\mid Z_0=z_0\sim\mathcal N(b_0z_0,\psi_0)$, then draw the partners independently as $X_j\sim\operatorname{Bernoulli}\{q_j(f)\}$. This produces an independent draw from \eqref{eq:condlaw} in $O(p)$ time.

The same construction extends directly to any $q$. The factor update becomes
\[
f\mid Z_0=z_0
\;\sim\;
\mathcal N_q\!\left(\mathbf b_0z_0,\ I_q-\mathbf b_0\mathbf b_0^{\!\top}\right),
\]
after which the $p-1$ partners remain conditionally independent. A Cholesky factor of the $q\times q$ conditional covariance costs $O(q^3)$ once, and each draw then costs $O(pq)$. Supplementary Note 2, Section 2.1 gives the full algorithm and proof.

The sequential construction above is an ancestral sampler. By drawing $Z_0$, then $f$, and then the partner alleles, it samples from $\Pr(X=x\mid X_0=s)$ without enumerating the $2^k$ partner configurations or storing their probabilities. This gives sampling its principal advantage: it can discover configurations differing at many partners, even when the factor-space partition is too large to enumerate. It also extends without a change in principle from one factor to several. Its advantage over uniform bit sampling is candidate-discovery efficiency, not a lower cost per draw. For a fixed configuration $x$ with HaploPerturb probability $\pi$, the expected waiting time is $1/\pi$ draws. Under uniform sampling over the partner alleles, the same configuration has probability $2^{-k}$ and expected waiting time $2^k$. Configurations that are probable under the allele frequencies and LD represented by HaploPerturb therefore recur sooner and rise naturally to the top of the frequency table.

The corresponding limitation is that candidate discovery is random. A finite run can miss a relevant configuration, especially when probability mass is spread across many nearly unique haplotypes, and repeated runs require a fixed seed and the same random-number-consumption order for exact reproducibility. Increasing the number of draws reduces but does not eliminate this risk. Sampling frequencies also contain Monte Carlo error, so they should not determine the final ordering when HaploPerturb probabilities can be evaluated directly.

For each lead state, we retained the 50 configurations with the largest sample counts. We then evaluated each retained configuration's HaploPerturb probability in \eqref{eq:condlaw} by Gauss--Legendre numerical integration: the latent-factor integral was approximated by a reproducible weighted sum at $1{,}024$ fixed nodes, with $2{,}048$ nodes used to refine the final ordering (Supplementary Note 2, Section 2.3). Thus sampling determines which candidates are considered, whereas deterministic probability evaluation determines their reported order. We used this hybrid strategy for all reported lists, which are based on $q=1$ fits. Candidate sampling extends directly to $q>1$; deterministic rescoring is the limiting step because product Gauss--Legendre quadrature scales as $O(pG^q)$.

\subsubsection{Deterministic factor-space enumeration}
\label{sec:closedform1}

The alternative strategy exploits the geometry of the fitted HaploPerturb factor model. Its principal advantages are reproducibility and economy: it uses no random draws, and at $q=1$ it reduces an exponential configuration space to a short, ordered path through the binary hypercube. Here each loading is a scalar $b_j$, $\psi_j=1-b_j^2$, and the conditionally most probable allele at factor value $f$ is
\begin{equation}
x_j^*(f)=\mathbf 1\{b_jf>\tau_j\},
\qquad
t_j=\tau_j/b_j\quad\text{when }b_j\ne0,
\label{eq:breakpoint-main}
\end{equation}
where $t_j$ is the partner's breakpoint. A partner with $b_j\ne0$ changes state exactly once as $f$ crosses $t_j$; a partner with $b_j=0$ is constant. Sorting the breakpoints partitions the line into at most $k+1$ intervals, and one interior point per interval visits every conditional mode. Enumerating these modes costs $O(p\log p)$ rather than $2^k$, before probability scoring.

For general $q$, the same coordinate-wise maximization at a fixed factor value gives
\begin{equation}
x^{*}(f)
=\operatorname*{argmax}_{x\in\{0,1\}^k}
\prod_{j=1}^{k}q_j(f)^{x_j}\{1-q_j(f)\}^{1-x_j},
\qquad
x_j^*(f)=\mathbf 1\{\mathbf b_j^{\!\top}f>\tau_j\}.
\label{eq:condmode}
\end{equation}
Sweeping the factor space therefore produces the finite candidate set
\begin{equation}
\mathcal C=\bigl\{x^*(f):f\in\mathbb R^q\bigr\},
\label{eq:candset}
\end{equation}
whose members are conditionally optimal somewhere in factor space. Conditional optimality at one $f$ does not, however, guarantee a high integrated probability in \eqref{eq:condlaw}. We therefore first score $\mathcal C$, let $\hat x=\operatorname*{argmax}_{x\in\mathcal C}\Pr(X=x\mid X_0=s)$, and expand the search to
\begin{equation}
\mathcal C^+
=\mathcal C\cup\bigl\{\hat x^{(j)}:j=1,\ldots,k\bigr\},
\qquad
\hat x^{(j)}=\hat x\text{ with coordinate }j\text{ flipped}.
\label{eq:candplus}
\end{equation}
At $q=1$, $\mathcal C$ traces a path through the binary hypercube and contains only two of the $k$ one-variant neighbors of an interior mode; adjoining every one-variant flip repairs this immediate omission. This exposes the strategy's central limitation: it searches configurations that are locally optimal somewhere in factor space, not all configurations having large integrated probability. A haplotype that is never a conditional mode may nevertheless accumulate substantial probability over a broad range of factor values. Supplementary Note 2, Section 2.2 gives the geometry and algorithms.

For $q>1$, the signs in \eqref{eq:condmode} change across the $k$ hyperplanes $H_j=\{f:\mathbf b_j^{\!\top}f=\tau_j\}$. Their regions enumerate $\mathcal C$, with
\begin{equation}
\lvert\mathcal C\rvert
\leq\sum_{i=0}^{q}\binom{k}{i}
=O(k^q)=O(p^q)
\qquad\text{for fixed }q.
\label{eq:cellcount}
\end{equation}
Thus the $q=1$ breakpoint sort generalizes to exact region enumeration whenever $p^q$ is small enough. Its favorable $O(p\log p)$ scaling at one factor does not persist with increasing $q$: the number of regions grows as $O(p^q)$, before their probabilities are scored. Moreover, the word ``exact'' applies only to enumerating the conditional modes. The expanded set $\mathcal C^+$ still excludes configurations that differ from $\hat x$ at two or more partners and therefore need not contain the global top $L$. Supplementary Note 2, Section 2.2 gives observed examples in which omitted configurations outrank members of $\mathcal C^+$.

\subsubsection{Choosing between the strategies}

Deterministic enumeration is attractive at $q=1$ and moderate $p$, where the breakpoint ordering is fast, transparent and exactly reproducible. It is useful for inspecting the fitted HaploPerturb model's geometry and for generating a compact candidate set without tuning a simulation size. It is less attractive when $q>1$ or $p$ is large, because the region count grows as $O(p^q)$, and it can miss high-integrated-probability configurations that are not conditional modes or their one-partner neighbors.

Model-guided sampling is preferable when the deterministic region count is large or when candidates separated by several partner flips must be discoverable. Its computational cost grows linearly in $p$ per draw and only modestly with $q$, but the number of draws required depends on how concentrated the conditional distribution is. If a haplotype has HaploPerturb probability $\pi$, the probability that $N$ independent draws miss it is $(1-\pi)^N$. The binomial union bound used in Section~\ref{sec:haplotype-probability} converts this fact into a prespecified probabilistic coverage check for the reported top-ten lists; a list that fails the check is reported only as a high-probability candidate list.

The two strategies are therefore complementary rather than interchangeable. Factor-space enumeration supplies a deterministic, interpretable search when its region count is manageable; conditional sampling supplies broader discovery when enumeration is restrictive or computationally prohibitive. Both still require probability evaluation to compare candidates on the target scale in \eqref{eq:condlaw}. Product quadrature with $G$ nodes per factor dimension costs $O(pG^q)$ for each retained candidate (Supplementary Note 2, Section 2.3), so rescoring ultimately limits both approaches at higher rank. Neither finite sampling nor the $\mathcal C^+$ search supplies an unconditional global top-$L$ guarantee. For the present analyses, we chose sampling for candidate discovery and fixed-node quadrature for final ranking because the same procedure applies across loci of widely varying size and admits an explicit coverage check.

\section{Model parameter estimation}\label{sec:estimation}
\label{sec:est}
 The object we want is the conditional law \eqref{eq:target}, whose top-ranked configurations are the haplotypes to be queried, and under the model of Section~\ref{sec:model} that law is a function of the parameters alone: given $(\mathbf{B},\tau)$ the residual variances follow from \eqref{eq:unitvar} as $\psi_j = 1 - \lVert\mathbf{b}_j\rVert^2$, the latent vector is $\mathcal{N}\bigl(0,\mathbf{B}\mathbf{B}^{\!\top} + \mathbf{D}\bigr)$, and thresholding at $\tau$ returns the law of $X$. Ranking haplotypes therefore requires nothing beyond $(\mathbf{B},\tau)$.

Our primary estimator fixes the marginal thresholds before estimating dependence. Let $m_j=\sum_{i=1}^{n}X_{ij}$ and use the Jeffreys-corrected frequency and corresponding threshold
\begin{equation}
\widetilde p_j=\frac{m_j+\tfrac12}{n+1},
\qquad
\widetilde\tau_j=\Phi^{-1}(1-\widetilde p_j).
\label{eq:fixed-margin}
\end{equation}
The correction keeps the threshold finite at rare variants; when $0<m_j<n$ its difference from the empirical frequency is $O(n^{-1})$. Under the working parameterization \eqref{eq:ogive}, fixing $\tau_j=\widetilde\tau_j$ makes the intercept a function of the loading,
\[
\beta_j(a_j)
=-\widetilde\tau_j\sqrt{1+\lVert a_j\rVert^2},
\qquad
\eta_j(f;a_j)
= a_j^{\!\top}f + \beta_j(a_j)
=a_j^{\!\top}f-\widetilde\tau_j\sqrt{1+\lVert a_j\rVert^2}.
\]
Let $A=(a_0,\ldots,a_k)^{\!\top}\in\mathbb R^{(k+1)\times q}$ denote the working-scale loading matrix for the $q$-factor model. For a prespecified uniqueness floor $\psi_{\min}>0$, put $A_{\min}=\sqrt{\psi_{\min}^{-1}-1}$. The fixed-margin marginal maximum-likelihood estimator is
\begin{equation}
\widehat A_q
=
\operatorname*{arg\,max}_{\lVert a_j\rVert\leq A_{\min},\,j=0,\dots,k}
\sum_{i=1}^{n}
\log\!\int_{\mathbb R^q}
\phi_q(f)
\prod_{j=0}^{k}
\Phi\!\left\{(2X_{ij}-1)\eta_j(f;a_j)\right\}
\,\mathrm df .
\label{eq:fixed-margin-mle}
\end{equation}
Throughout, a tilde denotes a fixed plug-in quantity computed from the marginal allele counts, whereas a hat denotes a quantity obtained from the fitted HaploPerturb model or from a subsequent optimization. The thresholds are not re-estimated in \eqref{eq:fixed-margin-mle}; consequently, their values in the fitted HaploPerturb model satisfy $\widehat\tau_j=\widetilde\tau_j$. The map following \eqref{eq:ogive} gives $\widehat\psi_j=(1+\lVert\widehat a_j\rVert^2)^{-1}$ and $\widehat{\mathbf b}_j=\widehat a_j\sqrt{\widehat\psi_j}$. Thus the HaploPerturb model has a valid correlation matrix and reproduces the estimated allele frequencies by construction, while estimating dependence under the Gaussian-threshold likelihood rather than by treating allelic correlations as latent correlations.

The integral in \eqref{eq:fixed-margin-mle} is concentrated differently for different observed haplotypes, especially when a uniqueness is near its floor. We therefore integrate around each haplotype's posterior factor mode. If
\[
h_i(f)
=-\tfrac12 f^{\!\top}f
+\sum_{j=0}^{k}
\log\Phi\!\left\{(2X_{ij}-1)\eta_j(f;a_j)\right\},
\]
write $\widehat f_i(A)=\operatorname*{arg\,max}_f h_i(f;A)$ and $H_i(A)=-\nabla^2h_i\{\widehat f_i(A);A\}$. At each likelihood-and-score evaluation for the reported $q=1$ fits, we collapsed identical observed haplotypes and retained their multiplicities. For each distinct pattern and the current loading parameters, we found the conditional factor mode $\widehat f_i(A)$ by Newton iteration with backtracking and computed its positive curvature $H_i(A)$. We then evaluated the pattern's likelihood contribution by one-dimensional posterior-adaptive Gauss--Hermite quadrature at nodes
\[
f_{im}=\widehat f_i(A)+\frac{u_m}{\sqrt{H_i(A)}},
\]
weighting that contribution by the pattern's multiplicity.

We maximized the resulting observed likelihood under the uniqueness constraint, using deterministic spectral starts and successively refined quadrature. The 1000 Genomes working correlation supplied only the starting values for the ROS/MAP optimization; the fixed margins and likelihood were computed entirely from ROS/MAP. We retained the converged start with the largest refined observed likelihood. Supplementary Note 1, Sections 1.1--1.3 specify the optimizer, starting vectors, quadrature sequence, tolerances and reporting checks.

\section{Constructing high-probability haplotypes for 1000 Genomes and ROS/MAP}
\label{sec:haplotype-analysis}

\subsection{Does one factor suffice?}
\label{sec:application-rank}

We first ask whether the one-factor model used by the conditional sampler and probability scorer accounts for most of the latent allelic variance. For a \(q\)-factor HaploPerturb fit with correlation \(\widehat{\mathbf R}_q=\widehat{\mathbf B}_q\widehat{\mathbf B}_q^{\!\top} +\widehat{\mathbf\Psi}_q\), define
\begin{equation}
\widehat{\mathrm{PVA}}(q)
=
\frac{\operatorname{tr}
  (\widehat{\mathbf B}_q\widehat{\mathbf B}_q^{\!\top})}
 {\operatorname{tr}(\widehat{\mathbf R}_q)}
=
\frac{1}{p}\sum_{j=1}^{p}\lVert\widehat{\mathbf b}_{j,q}\rVert^2
=
1-\frac{1}{p}\sum_{j=1}^{p}\widehat\psi_{j,q},
\label{eq:pva-application}
\end{equation}
where the last two equalities use the unit diagonal of \(\widehat{\mathbf R}_q\), and hence \(\operatorname{tr}(\widehat{\mathbf R}_q)=p\). This is the fraction of total latent variance assigned to the common factors.

\begin{figure}[t]
\centering
\includegraphics{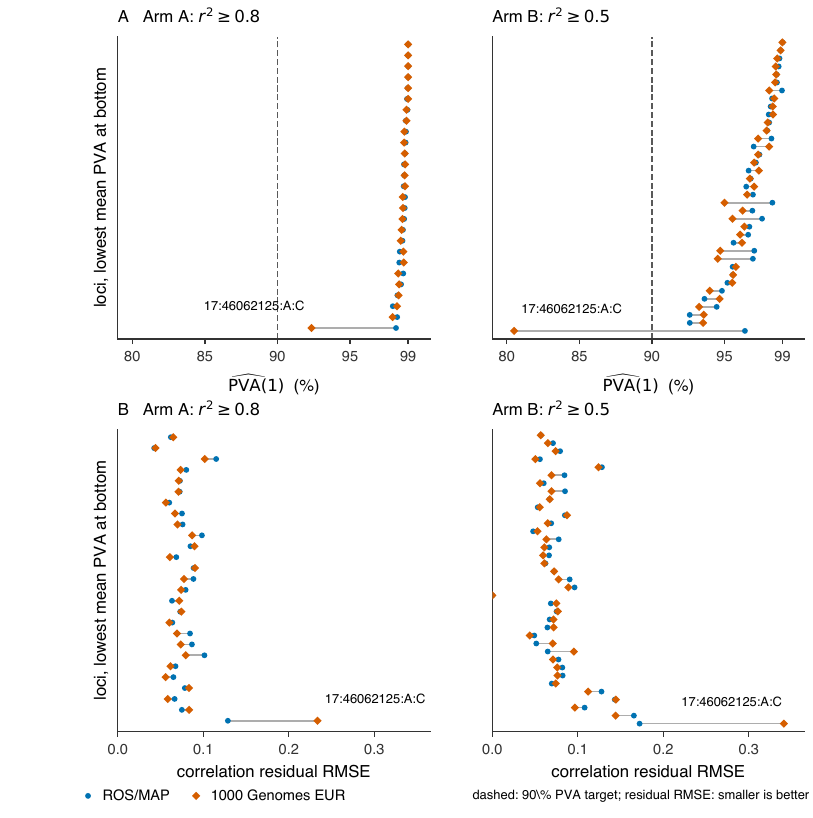}
\caption{\textbf{Adequacy of the one-factor HaploPerturb model across loci and cohorts.} Columns show Arm A ($r^2\geq0.8$) and Arm B ($r^2\geq0.5$). Each horizontal row represents one locus: blue circles show ROS/MAP, orange diamonds show the 1000 Genomes EUR panel, and a gray segment links the two cohort estimates. \textbf{(A)} Proportion of latent variance accounted for by the HaploPerturb factor fit, $\widehat{\mathrm{PVA}}(1)$. The vertical dashed line marks $90\%$. Values cannot exceed $99\%$ because the estimated uniqueness is constrained by $\psi_{\min}=0.01$. \textbf{(B)} Off-diagonal correlation-residual RMSE from equation~\eqref{eq:q1-correlation-rmse}, comparing the empirical correlations between phased binary allele indicators with the correlations implied by HaploPerturb. Smaller values indicate closer agreement. Within each arm, loci are ordered by their mean PVA across the two cohorts, from lowest at the bottom to highest at the top; the same rows are used in (A) and (B). Each cohort's statistics use its own phased haplotypes and the variants included in its HaploPerturb fit; partners absent from ROS/MAP are omitted from its comparison.}
\label{fig:q1-pva}
\end{figure}

We estimated HaploPerturb at \(q=1\) separately from the \(445\) ROS/MAP donors (\(890\) phased haplotypes) and the \(503\) unrelated 1000 Genomes EUR individuals (\(1{,}006\) phased haplotypes), under both partner-selection rules. In Arm A (\(r^2\geq0.8\), \(27\) loci), \(\widehat{\mathrm{PVA}}(1)\) ranged from \(97.95\%\) to \(99.00\%\) in ROS/MAP (median \(98.73\%\)) and from \(92.34\%\) to \(99.00\%\) in 1000 Genomes (median \(98.72\%\)). In Arm B (\(r^2\geq0.5\), \(37\) loci), it ranged from \(92.60\%\) to \(99.00\%\) in ROS/MAP (median \(97.01\%\)) and from \(80.50\%\) to \(98.99\%\) in 1000 Genomes (median \(96.74\%\)); see Figure~\ref{fig:q1-pva}. Arm A has higher PVA because its stricter LD threshold retains partners whose alleles track the lead more uniformly, making their shared dependence closer to a one-factor structure. Using the \(90\%\) reference threshold shown in Figure~\ref{fig:q1-pva}, \(127\) of the \(128\) cohort-by-arm locus fits exceeded the threshold; all \(128\) exceeded \(80\%\). The sole value below \(90\%\) was the largest Arm B locus, \texttt{17:46062125:A:C}, whose \(2{,}693\) partners produced \(\widehat{\mathrm{PVA}}(1)=80.50\%\) in 1000 Genomes.

Because a high PVA does not by itself establish off-diagonal fit, we also check the fit on the scale the construction consumes: the correlation between the binary alternate-allele indicators. Two \(p\times p\) correlation matrices are compared at each cohort--locus fit. The first, \(\widehat{\mathbf R}_{X}\), is empirical --- the Pearson correlation of the indicators \(X_j\) across that cohort's own phased haplotypes, over the \(p\) variants included in the fit; Supplementary Note 1, Section 1.4 gives its full definition. The second, \(\widehat{\mathbf R}_{X,1}\), is what the one-factor HaploPerturb model implies for those same indicators, computed from the HaploPerturb parameter estimates alone through the joint probability
\[
\Pr\bigl(X_j=1,\,X_\ell=1\bigr)
=
\Phi_2\bigl(-\widetilde\tau_j,\;-\widetilde\tau_\ell;\;
\widehat{\mathbf b}_j^{\!\top}\widehat{\mathbf b}_\ell\bigr),
\]
with \(\Phi_2(\cdot,\cdot;\rho)\) the standard bivariate normal upper-orthant probability at latent correlation \(\rho\), converted to a correlation between the two Bernoulli indicators using the fixed margins \(\Pr(X_j=1)=\Phi(-\widetilde\tau_j)=\widetilde p_j\). The same supplementary section gives the quadrature calculation and the limits on how the comparison may be read.

Both are correlation matrices, so both have unit diagonal and the whole discrepancy is off-diagonal:
\[
\widehat{\mathbf E}_{X}
=
\widehat{\mathbf R}_{X}-\widehat{\mathbf R}_{X,1},
\qquad
\operatorname{diag}\bigl(\widehat{\mathbf E}_{X}\bigr)=\mathbf 0 .
\]
Nor can it be marginal misfit, since the margins are held fixed at the empirical frequencies; it is pairwise dependence alone. We report the root mean square of the \(p(p-1)/2\) distinct entries of \(\widehat{\mathbf E}_{X}\),
\begin{equation}
\widehat{\operatorname{RMSE}}_{X}(1)
=
\left\{
\frac{2}{p(p-1)}
\sum_{j<\ell}
\widehat E_{X,j\ell}^{\,2}
\right\}^{1/2}
=
\frac{\bigl\lVert\widehat{\mathbf E}_{X}\bigr\rVert_{F}}
     {\sqrt{p(p-1)}},
\label{eq:q1-correlation-rmse}
\end{equation}
 The median of $\widehat{\operatorname{RMSE}}_{X}(1)$ was similar in all four analyses: \(0.075\) in ROS/MAP Arm A (range \(0.043\)--\(0.129\)), \(0.071\) in ROS/MAP Arm B (\(<0.001\)--\(0.172\)), \(0.072\) in 1000 Genomes Arm A (\(0.044\)--\(0.234\)), and \(0.072\) in 1000 Genomes Arm B (\(<0.001\)--\(0.341\)); see Figure~\ref{fig:q1-pva}B. For scale, the median fraction of squared off-diagonal correlation reproduced,
 \[1-\frac{\sum_{j<\ell}\widehat E_{X,j\ell}^{\,2}}{\sum_{j<\ell}\widehat R_{X,j\ell}^{2}},\]
 was \(99.31\%\)--\(99.44\%\) across the four cohort--arm groups. The group minima were \(98.33\%\) and \(95.04\%\) in ROS/MAP Arms A and B, and \(94.16\%\) and \(85.72\%\) in 1000 Genomes Arms A and B. The largest residual again occurred at \texttt{17:46062125:A:C} in 1000 Genomes Arm B. Supplementary Figure~1 shows its empirical, HaploPerturb-implied and residual correlation matrices: the positive residual blocks identify dependence that the one-factor approximation underestimates at this deliberately worst-fitting public-panel locus. Together, the PVA and residual checks support \(q=1\) as a fast working approximation for the present construction.

Low rank should not be conflated with sparse loadings. We therefore examined the estimated HaploPerturb latent-scale magnitudes \(\lvert\widehat b_j\rvert\) in the public 1000 Genomes analysis (Supplementary Figure~2). Using \(s_{\mathrm{eff}}=(\sum_j\widehat b_j^2)^2/\sum_j\widehat b_j^4\) as the effective number of loading rows, \(s_{\mathrm{eff}}/p\) had minimum/median \(0.994/1.000\) in Arm A and \(0.957/0.999\) in Arm B. Collecting \(90\%\) of \(\sum_j\widehat b_j^2\) required a median \(91.4\%\) and \(90.0\%\) of the variants, respectively, and the sorted magnitudes showed no separation into a near-zero group and an active group. Thus the one-factor HaploPerturb model is low rank but its factor is shared broadly across the selected variants; row sparsity is not supported by this diagnostic. This does not by itself prove that every sparse-PCA estimator will fit poorly, because that also depends on its target and tuning, but it removes empirical support for adopting sparse loadings as a modeling assumption here.

\subsection{Empirical validation of HaploPerturb rankings}
\label{sec:haplotype-probability}

To evaluate whether the HaploPerturb rankings identify haplotypes that occur in phased data, we examined two reported lists for every cohort, arm and locus: one conditional on the lead variant carrying its reference allele, \(X_0=0\), and one conditional on its alternate allele, \(X_0=1\). Each list contains up to ten configurations ordered by HaploPerturb probability; when fewer than ten configurations are possible after fixing the lead allele, the list is complete. Section~\ref{sec:closedform} and Supplementary Note 2, Sections 2.1, 2.3 and 2.4 describe how the lists were constructed, ranked and checked. All lists passed the prespecified coverage check except the alternate-conditioned 1000 Genomes Arm B list at \texttt{17:46062125:A:C}; results for that list therefore describe high-probability candidates rather than a certified global top ten.

We evaluated each generated haplotype against the phased data by its empirical conditional probability
\begin{equation}
\widehat P_{\mathrm{emp}}(x\mid X_0=s)
=
\frac{\sum_{i=1}^{n}\mathbf 1\{X_i=x,\ X_{i0}=s\}}
     {\sum_{i=1}^{n}\mathbf 1\{X_{i0}=s\}},
\qquad s\in\{0,1\}.
\label{eq:empirical-haplotype-probability}
\end{equation}
Here ``empirical probability'' means the observed frequency among the cohort's phased haplotypes that carry the same lead state used to generate the list: \(1{,}006\) total phased haplotypes in 1000 Genomes and \(890\) in ROS/MAP before conditioning.  In particular, a value of zero means that the exact generated haplotype was not observed in this finite panel; it does not establish that the haplotype has zero population probability. This limitation becomes consequential as the number of variants grows, because exact high-dimensional haplotypes are sparsely repeated.

\begin{figure}[htbp]
\centering
\includegraphics[width=0.70\textwidth]{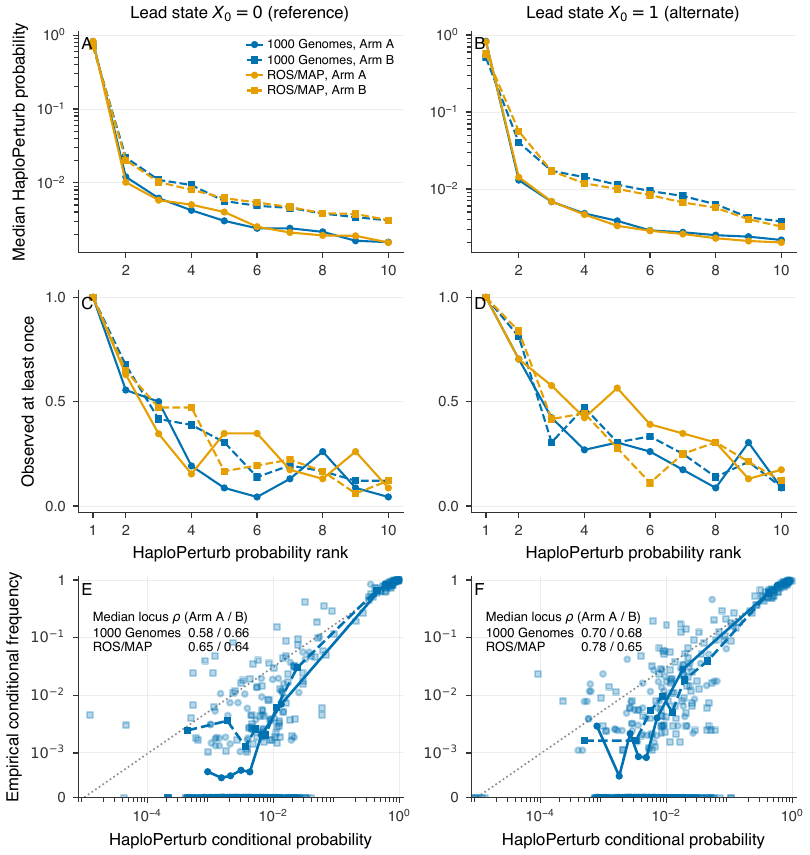}
\caption{\textbf{HaploPerturb probability ranks correspond to empirical haplotype frequencies.} The left column conditions on the reference lead state ($X_0=0$); the right column conditions on the alternate state ($X_0=1$). Blue denotes 1000 Genomes and orange denotes ROS/MAP; circles with solid lines denote Arm A ($r^2\geq0.8$), and squares with dashed lines denote Arm B ($r^2\geq0.5$). \textbf{(A,B)} Median HaploPerturb conditional probability at ranks one through ten. The steep drop after rank one indicates that HaploPerturb probability is concentrated in the leading configuration. \textbf{(C,D)} Proportion of loci at which the candidate at each HaploPerturb rank occurs at least once among phased haplotypes carrying the corresponding lead allele. The rank-one candidate is observed at every locus, with observation rates generally decreasing at lower ranks. \textbf{(E,F)} HaploPerturb conditional probability versus empirical conditional frequency for individual 1000 Genomes locus--candidate pairs. Solid and dashed blue curves connect mean empirical frequencies within eight equal-count bins of HaploPerturb probability; the gray dotted line marks equality. Text within each panel gives the median within-locus Spearman correlation for Arm A and Arm B in both cohorts. The symmetric-logarithmic vertical scale retains candidates with zero observed frequency while separating small positive frequencies. Candidate-level ROS/MAP frequencies are not displayed because the underlying donor genotypes are access-controlled; their rank summaries appear in (A--D) and their correlation summaries appear in (E,F). An empirical frequency of zero means that the candidate was not observed in the finite cohort, not that its population probability is zero.}
\label{fig:top10-rank}
\end{figure}

The HaploPerturb rank-one haplotype was observed at every locus for both lead states. Under \(X_0=0\), its median empirical conditional probability was \(0.955\) and \(0.802\) in 1000 Genomes Arms A and B and \(0.967\) and \(0.802\) in ROS/MAP Arms A and B. Under \(X_0=1\), the corresponding medians were \(0.931\), \(0.668\), \(0.936\), and \(0.656\). Observation became less frequent down both HaploPerturb lists. At rank two, the reference-conditioned candidate was observed at \(55.6\%\) and \(67.6\%\) of 1000 Genomes Arm A and B loci and at \(63.0\%\) and \(64.9\%\) of ROS/MAP Arm A and B loci; the alternate-conditioned fractions were \(70.4\%\), \(81.1\%\), \(70.4\%\), and \(83.8\%\), respectively. By rank ten all eight fractions were between \(4.3\%\) and \(17.4\%\) (Figure~\ref{fig:top10-rank}C--D). Although the individual lower-ranked haplotypes were rarely observed, their cumulative empirical coverage increased monotonically. The top ten reference-conditioned haplotypes covered a median \(0.976\) and \(0.928\) of the empirical conditional mass in 1000 Genomes Arms A and B and \(0.979\) and \(0.934\) in ROS/MAP. Under \(X_0=1\), the corresponding median coverages were \(0.959\), \(0.907\), \(0.973\), and \(0.918\).

The ordering nevertheless contains information. For \(X_0=0\), the median within-locus Spearman correlation between HaploPerturb probability and empirical frequency was \(0.580\) and \(0.657\) in 1000 Genomes Arms A and B and \(0.651\) and \(0.636\) in ROS/MAP. For \(X_0=1\), the medians were \(0.705\), \(0.679\), \(0.777\), and \(0.649\). All \(256\) cohort-by-arm-by-lead-state locus correlations were positive. The candidate-level 1000 Genomes points and their binned trends show the same positive association directly (Figure~\ref{fig:top10-rank}E--F), while retaining the many exact zeros produced by finite-panel sampling. Results were consistent in both cohorts: haplotypes assigned higher HaploPerturb ranks were generally observed more often. Many exact haplotypes below rank two were not observed in either cohort, as expected when a limited number of chromosomes is sampled from a large set of possible haplotypes. The finite cohort data thus support the HaploPerturb ordering, but provide too little information to assess the accuracy of the HaploPerturb probabilities for lower-ranked haplotypes.

\subsection{Cross-cohort portability of HaploPerturb constructions}
\label{sec:cohort-haplotype-agreement}
To test whether a public reference panel could supply the sequence backgrounds without donor-specific genotypes, we compared the independently estimated 1000 Genomes and ROS/MAP HaploPerturb lists at every matched arm, locus, and lead state. No model was refitted for this comparison. Let \(P_\ell^{(c)}\) denote the partner coordinates modeled at locus \(\ell\) in cohort \(c\in\{\mathrm{1KG},\mathrm{RM}\}\). Because 1000 Genomes defined the requested partner set and ROS/MAP omitted unavailable variants, \(P_\ell^{(\mathrm{RM})}\subseteq P_\ell^{(\mathrm{1KG})}\). We first separated loci with identical partner sets from loci with fewer partners in ROS/MAP. At loci with identical sets, we compared haplotypes over all modeled partners. At loci with different sets, we projected both haplotypes onto the shared coordinates \(C_\ell=P_\ell^{(\mathrm{1KG})}\cap P_\ell^{(\mathrm{RM})}=P_\ell^{(\mathrm{RM})}\) before comparing them. Thus missingness in ROS/MAP was not counted as allelic agreement or disagreement.

Let \(x_{\ell srj}^{(c)}\) be the allele assigned to partner \(j\) at HaploPerturb rank \(r\), conditional on lead state \(s\). For corresponding ranks, we computed the Hamming distance and its partner-normalized form
\begin{equation}
d_{\ell sr}=\sum_{j\in C_\ell}
\mathbf 1\!\left\{x_{\ell srj}^{(\mathrm{1KG})}
\ne x_{\ell srj}^{(\mathrm{RM})}\right\},
\qquad
\delta_{\ell sr}=\frac{d_{\ell sr}}{|C_\ell|}.
\label{eq:cohort-haplotype-agreement}
\end{equation}
Rank one was the primary comparison. For the top-five and top-ten lists, we also computed the exact intersection divided by the size of the smaller list after projection. To describe near matches without penalizing a change in rank, each configuration was matched to its nearest configuration in the other cohort by Hamming distance, in both directions. These comparisons use only HaploPerturb sequence configurations and do not use the Fujita eQTL labels.

\begin{table}[htbp]
\centering
\caption{\textbf{Agreement of HaploPerturb high-probability haplotypes estimated from 1000 Genomes and ROS/MAP.} A rank-one pair compares the two cohorts at one locus and lead state. When partner sets differ, all identity and distance calculations use only shared coordinates. Missing-partner entries give the median and range among loci with different sets. The top-ten shared fraction is the median across locus--state comparisons of the exact intersection divided by the smaller projected list size. The final column summarizes the pooled bidirectional Hamming distances to the nearest configuration in the other cohort as median [95th percentile, maximum].}
\label{tab:cohort-haplotype-agreement}
\footnotesize
\setlength{\tabcolsep}{3.5pt}
\begin{tabular}{llccccc}
\toprule
Arm & Partner sets & Loci & Missing partners & Rank one exact & Top ten shared & Top-ten nearest \(d\) \\
\midrule
A & Same & 22 & 0 & \(44/44\) & 0.85 & \(0\ [1,37]\) \\
A & Different; shared only & 5 & \(1\ [1,40]\) & \(10/10\) & 0.79 & \(0\ [1,13]\) \\
B & Same & 25 & 0 & \(50/50\) & 0.90 & \(0\ [1,75]\) \\
B & Different; shared only & 12 & \(1\ [1,132]\) & \(23/24\) & 0.84 & \(0\ [1,2{,}560]\) \\
\bottomrule
\end{tabular}
\end{table}

Because the partner set does not depend on the conditioned lead state, partner availability is counted once per arm--locus. The partner sets were identical at \(22/27\) Arm A loci and \(25/37\) Arm B loci (Table~\ref{tab:cohort-haplotype-agreement}). Among the remaining loci, ROS/MAP lacked \(45\) Arm A partners across five loci and \(152\) Arm B partners across 12 loci. The median number missing was one in both arms, with ranges of 1--40 and 1--132, respectively.

When the partner sets were identical, all \(44/44\) Arm A and \(50/50\) Arm B rank-one haplotype pairs were identical. After projection onto shared partners at loci with different sets, all \(10/10\) Arm A pairs and \(23/24\) Arm B pairs were identical. Equivalently, both members of the rank-one contrast agreed at all \(27\) Arm A loci and at \(36/37\) Arm B loci. The only exception was the alternate-state comparison at \texttt{17:46062125:A:C}: all \(2{,}561\) shared partners differed. This is the previously identified diffuse Arm B locus whose 1000 Genomes alternate-state list was not certified as a global top ten, so the exception cannot be separated from instability in that candidate search.

The lower-ranked lists were less often identical, but generally contained the same or neighboring configurations. Across locus--state comparisons, the median exact shared fraction of the top-ten lists was \(0.85\) and \(0.90\) at Arm A and B loci with identical partner sets, and \(0.79\) and \(0.84\) after projection at loci with different sets. In all four strata, the median nearest-configuration Hamming distance was zero and its 95th percentile was one partner. The corresponding 95th percentiles of the normalized distance \(\delta\) were \(0.056\), \(0.080\), \(0.033\), and \(0.067\), respectively. The tails were much wider, however, particularly at the diffuse chromosome 17 locus (Table~\ref{tab:cohort-haplotype-agreement}). Thus the leading HaploPerturb sequence contrast was highly portable on coordinates observed in both cohorts, whereas the ordering and composition of lower-ranked candidates were less stable. This agreement alone does not establish equal AlphaGenome effects or eQTL enrichment, which will be explored in Section~\ref{sec:external-reference-eqtl}.

\subsection{Two simple strategies for assigning partner alleles and their performance on 1000 Genome data}
\label{sec:simple-haplotype-baselines}

We compare two simple strategies, \emph{empirical conditional-mode} and \emph{LD-sign}, with our proposed \emph{HaploPerturb rank-one} strategy.

The \emph{empirical conditional-mode} strategy uses the observed phased haplotypes directly. It retains only lead allele to either 0 or 1, counts how often each
complete combination of partner alleles occurs, and then chooses the most frequent combination. If several combinations have the same maximum count, a fixed coordinate-order convention selects one of them.

The \emph{LD-sign} strategy examines each partner variant separately. If the lead and partner alternate-allele indicators are positively correlated, the partner is assigned the same reference-or-alternate state as the specified lead allele; if they are negatively correlated, the partner is assigned the opposite state. If the
correlation is exactly zero, the strategy uses the more frequent partner allele among panel
haplotypes carrying the specified lead allele, choosing the reference allele if those counts are
also tied.

Both simpler strategies equaled the HaploPerturb rank-one configuration in all $54/54$ Arm A
locus-state comparisons and in $73/74$ Arm B comparisons. There were no tied empirical modes and no zero-correlation fallbacks. The sole exception was the alternate-conditioned Arm B comparison at
\texttt{17:46062125:A:C}. There the empirical-mode and LD-sign configurations were identical to one
another and to HaploPerturb rank two; the empirical configuration occurred on 6 of the 256 phased
haplotypes carrying the alternate lead allele. Both differed from the reported HaploPerturb rank-one
candidate at all 2,693 partners.

\subsection{Systematic evaluation of when the three strategies differ}
\label{sec:ranking-simulation}

To determine systematically whether and when the three strategies select different partner-allele
configurations, we constructed simulations in which the population haplotype distribution was
known. For each setting, we identified the true most-probable partner-allele configuration
conditional on the lead allele and asked whether it was recovered by HaploPerturb, the empirical
conditional-mode and LD-sign strategies. We varied four features expected to affect their
performance: allele frequency, dependence structure, number of partners and reference-panel size.

\begin{figure}[htbp]
\centering
\includegraphics[width=\linewidth]{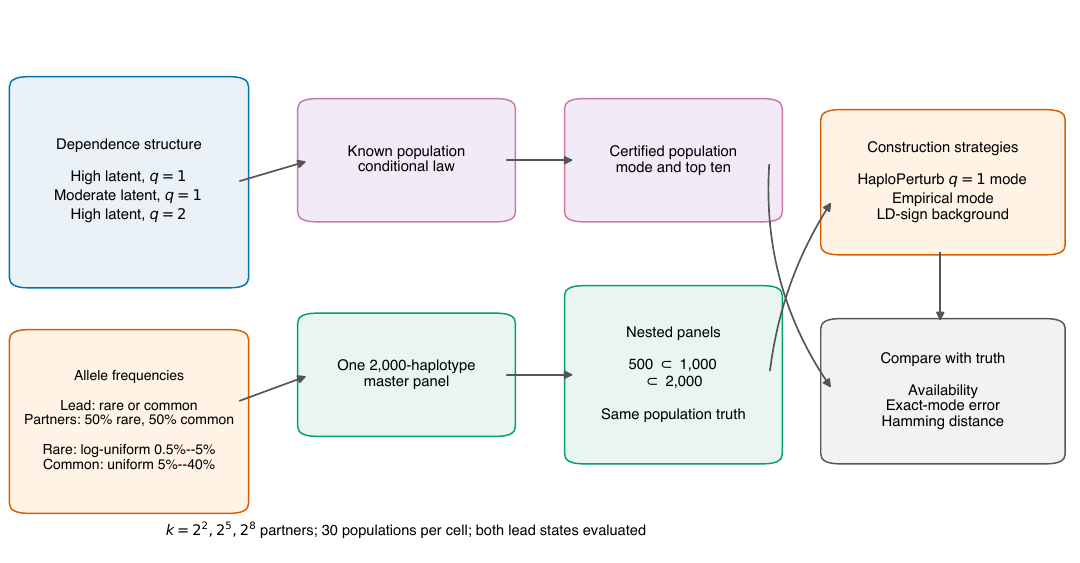}
\caption{\textbf{Design of the known-truth finite-panel simulation.} The design varies lead-allele
frequency, partner-allele frequencies, dependence structure, partner count and panel size. Each
population supplies one known conditional law and one 2,000-haplotype master panel; the 500- and
1,000-haplotype panels are nested prefixes. The HaploPerturb mode, empirical conditional mode and LD-sign
configuration are compared with the certified population mode for both lead states.}
\label{fig:ranking-simulation-design}
\end{figure}

The design crossed two lead-frequency classes, three data-generating dependence settings, three
partner counts and three panel sizes (Figure~\ref{fig:ranking-simulation-design}). Rare lead frequencies were drawn
log-uniformly from $0.5\%$ to $5\%$, whereas common lead frequencies were drawn uniformly from
$5\%$ to $40\%$. In every population, half of the partners were rare under the same log-uniform
distribution and half were common under the same uniform distribution. The partner counts were
$2^2$, $2^5$ and $2^8$, and the panel sizes were 500, 1,000 and 2,000 phased haplotypes. The three
dependence settings were high latent dependence generated by a one-factor model, moderate latent
dependence generated by a one-factor model, and high latent dependence generated by a two-factor
model but analyzed with the production one-factor model. These labels refer to latent dependence:
with mixed allele frequencies, high latent correlation need not produce uniformly high observed
$r^2$.

For each lead-frequency, dependence and partner-count setting, we generated 30 population
distributions. From each population, we simulated one master panel of 2,000 haplotypes and used its
first 500 and first 1,000 haplotypes as the two smaller panels. Thus all three panel sizes were drawn
from the same population, and increasing the panel size added information without changing the true
most-probable configuration. We estimated the one-factor HaploPerturb model separately in each panel and
applied all three selection strategies for both lead states, giving 1,620 HaploPerturb fits and 3,240
lead-state comparisons. The main comparison focuses on $X_0=1$, where sampling a rare alternate
allele is most difficult. If a panel contained no haplotype with $X_0=1$, the empirical
conditional-mode strategy was recorded as unavailable; it was not replaced by a fallback or counted
as an error. We therefore report unavailability separately from recovery error among panels in which
the strategy was available.

The certified search used to obtain the population and HaploPerturb top-ten lists was tested at five
partner counts: $k=4,8,12,32,256$. At $k=4,8,12$, its lists matched complete enumeration for every
combination of the three partner counts, three data-generating dependence settings and two lead
states ($3\times3\times2=18$ tests). At $k=32,256$, complete enumeration was infeasible. For every
combination of the two partner counts, three dependence settings and two lead states
($2\times3\times2=12$ tests), the search ended only after every remaining branch had either been
evaluated or shown by its probability upper bound to be unable to reach the current top-ten cutoff.
Thus no unevaluated configuration could enter the returned top ten under the quadrature
approximation. In the reported simulation, all 3,240 population searches and all 3,240 HaploPerturb
searches produced certified lists; all 2,160 four-partner population and HaploPerturb searches also passed
their exhaustive checks, including two cases with non-unique ordering at the top-ten cutoff.
Supplementary Note~4 gives the complete construction and search-validation details.

The three strategies separated most clearly when the alternate lead allele was rare. The empirical
conditional-mode strategy was usually available: only one of the 270 rare-lead populations had no
alternate-bearing haplotype at $n=500$ and $n=1{,}000$, and every population had at least one by
$n=2{,}000$ (Supplementary Figure~3). Its main failure was instead that the complete partner
configuration was too sparsely repeated to reveal the population mode. For rare leads, empirical-mode
error declined only from $76.7\%$ to $43.3\%$ with 32 partners and from $93.3\%$ to $74.4\%$ with
256 partners as the panel grew from 500 to 2,000. At 256 partners, empirical modes were commonly tied
and remained far from the population mode in Hamming distance (Supplementary Figure~4).

The LD-sign strategy also failed in many rare-lead settings, but for a different reason. It requires
an estimable lead--partner correlation for every partner. At $n=500$, all required signs were
available in only $78/90$ rare-lead panels with 32 partners and $46/90$ with 256 partners because
some sampled variants were monomorphic. Among evaluable 32-partner panels, its error was $55.1\%$,
$52.8\%$ and $53.3\%$ at $n=500$, 1,000 and 2,000, respectively. The strategy sometimes succeeded
in a favorable setting, but a pairwise sign does not use the allele-frequency margins and does not
ensure that the separately assigned partner alleles form the most probable joint configuration.

HaploPerturb avoided both of these finite-panel failure modes: it neither required an
exact complete configuration to recur nor relied on every sample correlation sign being defined. It
was available in all rare-lead panels. Its error declined from $30.0\%$ to $14.4\%$ with 32 partners
and from $31.1\%$ to $1.1\%$ with 256 partners as the panel grew from 500 to 2,000. At $n=2{,}000$,
these errors were substantially below the corresponding empirical-mode errors of $43.3\%$ and
$74.4\%$. Thus HaploPerturb provided the most stable and reliable construction across the
high-dimensional rare-lead settings.

\begin{figure}[htbp]
\centering
\includegraphics[width=\linewidth]{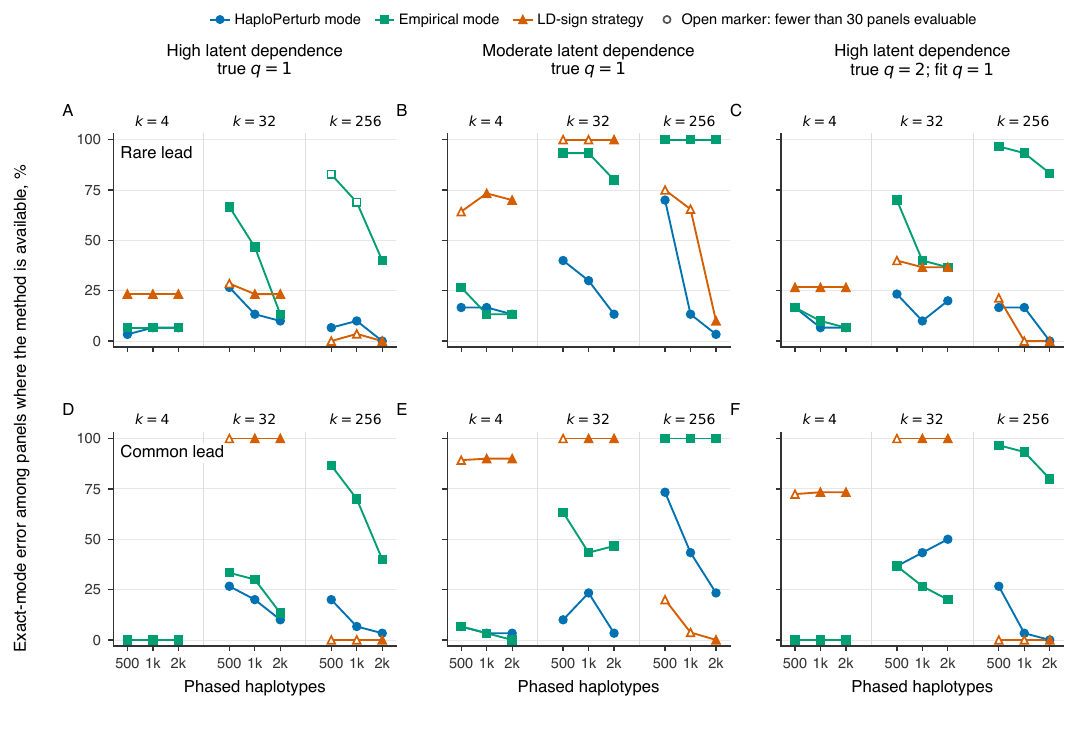}
\caption{\textbf{Conditional error for alternate-state haplotype construction.}
\textbf{(A--C)} Exact population-mode error for rare leads and \textbf{(D--F)} for common leads,
conditional on the displayed method being available. Columns show the three dependence settings;
within each panel, groups show 4, 32 and 256 partners and points show 500, 1,000 and 2,000 phased
haplotypes. Each point represents 30 populations. An open marker indicates that fewer than 30 were
evaluable for that method; no fallback result is included. Empirical-mode unavailability caused by
the absence of alternate-lead carriers is shown in Supplementary Figure~3.}
\label{fig:ranking-simulation-results}
\end{figure}

The same high-dimensional sparsity affected empirical modes for common leads: at 256 partners their
error declined only from $94.4\%$ to $73.3\%$, whereas HaploPerturb-mode error declined from $40.0\%$ to
$8.9\%$. With only four partners, however, the HaploPerturb and empirical modes were usually similarly
accurate. HaploPerturb also depends on an adequate low-rank representation.

Under deliberately misspecified two-factor truth with a common lead and 32 partners, HaploPerturb-mode error was $36.7\%$, $43.3\%$ and $50.0\%$ at the three panel sizes. More data cannot remove bias from an inadequate factor structure, so residual-correlation diagnostics remain necessary (Supplementary Figure~5). The simulation therefore supports HaploPerturb when complete haplotypes are sparse, especially for rare leads, while also motivating model-adequacy checks before treating its first-ranked configuration as definitive.

\section{Cell-type-specific eQTL recovery for Alzheimer's disease}
\label{sec:application}

\subsection{ROS/MAP donor-derived haplotype analysis}
\label{sec:internal-donor-eqtl}

We next ask whether the constructed contrasts recover cell-type-specific eQTLs. For each arm, locus $\ell$, gene $g$, AlphaGenome biosample $c$, HaploPerturb rank $r$, and lead state $s$, let $\mathcal E_g$ be the union of the gene's annotated exonic bases and let $Y_{\ell gcr}^{(s)}$ be the mean predicted RNA signal over $\mathcal E_g$. Overlapping exons are counted once, introns are excluded, and a gene is retained only when its complete exon union lies inside the 1.05-Mb AlphaGenome input window. Full transcript counts are called by AlphaGenome and reported there; the 1,000-bp terminal-exon summary is a sensitivity analysis reported in Supplementary Figure~6. The cell types with a defensible mapping to the experimental benchmark are astrocytes (AlphaGenome astrocyte), excitatory neurons (glutamatergic neuron), and microglia (CD14-positive monocyte, explicitly a myeloid proxy). We do not aggregate across these biosamples. The rank-one and average-top-$L$ effects, for $L\in\{5,10\}$, are
\begin{align}
D_{\ell gc}^{(1)}
&=\log\{1+Y_{\ell gc1}^{(1)}\}-\log\{1+Y_{\ell gc1}^{(0)}\},\\
D_{\ell gc}^{(L)}
&=\log\!\left\{1+R_{\ell1L}^{-1}\sum_{r=1}^{R_{\ell1L}}Y_{\ell gcr}^{(1)}\right\}
 -\log\!\left\{1+R_{\ell0L}^{-1}\sum_{r=1}^{R_{\ell0L}}Y_{\ell gcr}^{(0)}\right\},
\label{eq:eqtl-contrasts}
\end{align}
where $R_{\ell sL}=\min(L,\text{number available})$. We report rank one, average top five, and average top ten separately.

\paragraph*{Single-lead-variant comparator}
To separate the value of a HaploPerturb background from the value of changing the GWAS lead alone, we make a second, sequence-level comparison using the same AlphaGenome version, output tracks, 1.05-Mb input windows and complete-exon-union summaries. Let $G_\ell$ be the GRCh38 sequence in the window centered at lead $\ell$, whose reference and alternate alleles are $a_{\ell0}$ and $a_{\ell1}$. The single-lead sequences are
\begin{equation}
S_{\ell}^{\mathrm{SL},0}=G_\ell,
\qquad
S_{\ell}^{\mathrm{SL},1}
=G_\ell\text{ with only }a_{\ell0}\text{ replaced by }a_{\ell1}
\text{ at the lead position}.
\label{eq:single-lead-sequences}
\end{equation}
Every partner allele therefore remains GRCh38 in both sequences. No ROS/MAP genotype or HaploPerturb partner configuration enters this comparator. A lead appearing in both arms is predicted only once; the Arm A and Arm B labels are attached afterward according to whether that lead's window passes the corresponding partner-selection rule. If $Y_{\ell gc}^{\mathrm{SL},s}$ is the resulting complete-exon-union RNA prediction, its effect and ranking score are
\begin{equation}
D_{\ell gc}^{\mathrm{SL}}
=\log\{1+Y_{\ell gc}^{\mathrm{SL},1}\}
 -\log\{1+Y_{\ell gc}^{\mathrm{SL},0}\},
\qquad
T_{\ell gc}^{\mathrm{SL}}=|D_{\ell gc}^{\mathrm{SL}}|.
\label{eq:single-lead-score}
\end{equation}
The HaploPerturb scores are defined on the same scale, $T_{\ell gc}^{(L)}=|D_{\ell gc}^{(L)}|$ for $L\in\{1,5,10\}$. Thus the strategy comparison changes the input sequence but not the predictor, RNA track, gene summary or effect transformation.

For each arm and cell type, the comparison universe $\mathcal U$ is the intersection of locus--gene pairs covered by both AlphaGenome strategies and genes tested by Fujita et al. A pair is positive when Fujita et al. report a two-step-FDR-significant variant--gene association whose variant lies in the same AlphaGenome input window. Write $M=|\mathcal U|$, $Z_{\ell g}\in\{0,1\}$ for this truth indicator, and $K=\sum_{(\ell,g)\in\mathcal U}Z_{\ell g}$. For strategy $h\in\{\mathrm{top1},\mathrm{top5},\mathrm{top10},\mathrm{SL}\}$, let $\mathcal C_h$ be the $K$ pairs with largest $T_{\ell gc}^h$. The matched-count enrichment is
\begin{equation}
\widehat{\operatorname{Enr}}_h
=\frac{K^{-1}\sum_{(\ell,g)\in\mathcal C_h}Z_{\ell g}}{K/M}
=\frac{M}{K^2}\sum_{(\ell,g)\in\mathcal C_h}Z_{\ell g}.
\label{eq:strategy-enrichment}
\end{equation}
We resample whole lead windows with replacement 10,000 times and recompute $M$, $K$, the top-$K$ sets and all four enrichments in every draw. We report the 2.5th and 97.5th percentiles of each bootstrap distribution as a 95\% whole-locus bootstrap percentile interval. The same sampled windows are used for every strategy, so the bootstrap distribution of $\widehat{\operatorname{Enr}}_{h}-\widehat{\operatorname{Enr}}_{h'}$ is paired and preserves within-window dependence in both scores and truth. The main-text comparison is prespecified for the microglial proxy. Supplementary Note~5 gives the resampling algorithm and seeds.

AlphaGenome is deterministic, so its contrast has no sampling standard error. Our inferential unit is instead the lead window: the whole-locus bootstrap in equation~\eqref{eq:strategy-enrichment} quantifies how the enrichment changes across the available loci.

\begin{figure}[htbp]
\centering
\includegraphics[width=0.62\textwidth]{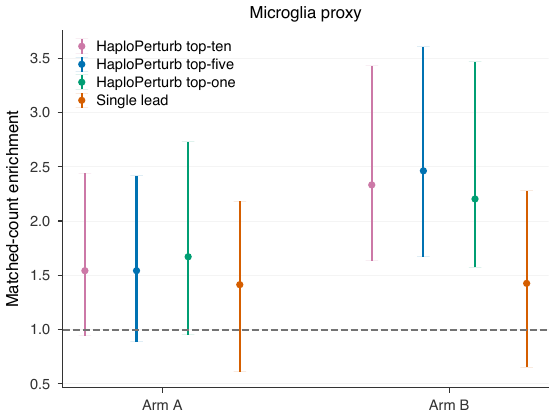}
\caption{\textbf{Microglial eQTL enrichment from HaploPerturb and single-lead perturbations.} Arm A uses partners with $r^2\geq0.8$ and Arm B uses $r^2\geq0.5$. HaploPerturb top one, top five and top ten summarize AlphaGenome contrasts between high-probability haplotypes carrying opposite lead alleles; the single-lead strategy changes only the lead allele on GRCh38. For each strategy, the $K$ highest-scoring locus--gene pairs are selected, where $K$ equals the number of Fujita two-step-FDR-positive pairs in the common evaluation set. Matched-count enrichment is the positive fraction among those selected pairs divided by the positive fraction in the full evaluation set, so the dashed line at one denotes no enrichment. Points are estimates based on complete gene-level exon unions, and error bars are the $2.5$th and $97.5$th percentiles from $10{,}000$ whole-locus bootstrap samples. All Arm A intervals cross one. In Arm B, the intervals for all three HaploPerturb summaries lie above one, whereas the single-lead interval crosses one.}
\label{fig:eqtl-strategy}
\end{figure}

The complete-transcript benchmark identifies a microglial ranking signal most clearly in Arm B (Figure~\ref{fig:eqtl-strategy}). In Arm A, the common universe contains $M=176$ locus--gene pairs, of which $K=37$ are Fujita-positive. Single-lead, HaploPerturb-top-one, HaploPerturb-top-five and HaploPerturb-top-ten rankings recover 11, 13, 12 and 12 positives, respectively. The estimated enrichments are $1.41$, $1.67$, $1.54$ and $1.54$; their corresponding 95\% whole-locus bootstrap percentile intervals are $[0.61,2.19]$, $[0.95,2.73]$, $[0.89,2.42]$ and $[0.95,2.44]$. Each enrichment interval includes the null value of one, and each 95\% paired bootstrap interval for the difference between two strategies includes the null value of zero.

In Arm B, $M=251$ and $K=44$. Single lead recovers 11 positives, for enrichment $1.43$ $[0.65,2.28]$; HaploPerturb top one recovers 17, for $2.20$ $[1.57,3.47]$; HaploPerturb top five recovers 19, for $2.46$ $[1.67,3.61]$; and HaploPerturb top ten recovers 18, for $2.33$ $[1.63,3.43]$. All three HaploPerturb intervals exclude one, whereas the single-lead interval includes one. Moreover, the paired HaploPerturb-minus-single-lead differences exclude zero: $0.78$ $[0.22,1.88]$ for top one, $1.04$ $[0.27,2.07]$ for top five and $0.91$ $[0.16,1.95]$ for top ten. Differences among the three HaploPerturb summaries all include zero; in particular, top-ten minus top-five is $-0.13$ $[-0.54,0.34]$. Thus Arm B supports enrichment attributable to HaploPerturb backgrounds but does not distinguish how many leading HaploPerturb configurations should be averaged.

Complete-transcript and terminal-exon results across all three cell types are reported in Supplementary Figure~6. In the microglial analysis, restricting the AlphaGenome scores to the terminal-exon region produced the same general pattern as the complete-transcript analysis: HaploPerturb strategies tended to have higher enrichment than the single-lead strategy. However, every 95\% paired bootstrap interval for a HaploPerturb-minus-single-lead difference included zero, so the terminal-exon analysis does not provide clear evidence that the HaploPerturb strategies outperform the single-lead strategy. The complete-transcript summary provided stronger evidence in this dataset, but the comparison does not establish that it is generally more powerful. In addition,
neither the astrocyte nor excitatory-neuron analysis shows positive enrichment: no 95\% interval lies wholly above one under either region definition. This lack of enrichment may reflect limited AlphaGenome accuracy for these cell types, but it could also arise from imperfect correspondence between the available AlphaGenome tracks and the Fujita cell types, limited eQTL power, or the chosen RNA-summary regions.

\subsection{1000 Genomes reference-panel haplotype analysis}
\label{sec:external-reference-eqtl}

We next ask whether donor-specific whole-genome sequencing is necessary for the enrichment result. We replace the ROS/MAP-derived HaploPerturb backgrounds in Section~\ref{sec:internal-donor-eqtl} with HaploPerturb backgrounds estimated from the 503-person 1000 Genomes EUR panel in Section~\ref{sec:haplotype-probability}. For every arm, locus, lead state and HaploPerturb rank, the public-panel configuration is written onto the same 1.05-Mb GRCh38 window and submitted to the same AlphaGenome model. We then use the same RNA tracks, complete gene-level exon unions and log-one-plus contrast in equation~\eqref{eq:eqtl-contrasts}. The resulting public-panel scores \(T_{\ell gc}^{(L,\mathrm{1KG})}\), \(L\in\{1,5,10\}\), therefore differ from the donor-panel scores only in the panel used to estimate the HaploPerturb backgrounds. When fewer than \(L\) configurations are available under either lead state, both analyses average the complete smaller paired support.

The comparison is restricted to the intersection of locus--gene pairs covered by all six HaploPerturb panel-by-rank strategies---ROS/MAP and 1000 Genomes top one, top five and top ten---the single-lead strategy, and the Fujita tested set. The Fujita two-step-FDR indicator and the definition of a positive locus--gene pair are unchanged from Section~\ref{sec:internal-donor-eqtl}. Within each arm and cell type, every strategy calls the top \(K\) scores in this common universe, where \(K\) is the number of Fujita-positive pairs. We resample whole lead windows 10,000 times and use the same sampled windows for both panels and all ranks. The direct comparison at rank summary \(L\) is
\begin{equation}
\widehat\Delta_c^{(L)}
=
\widehat{\operatorname{Enr}}_{c,\mathrm{1KG}}^{(L)}
-
\widehat{\operatorname{Enr}}_{c,\mathrm{RM}}^{(L)},
\qquad L\in\{1,5,10\},
\label{eq:reference-panel-enrichment-difference}
\end{equation}
with percentile intervals taken from the paired bootstrap differences. This design asks whether the public reference panel preserves the ROS/MAP eQTL-ranking enrichment, not whether the two panel-specific HaploPerturb models assign identical probabilities to each haplotype.

\begin{figure}[t]
\centering
\includegraphics[width=\linewidth]{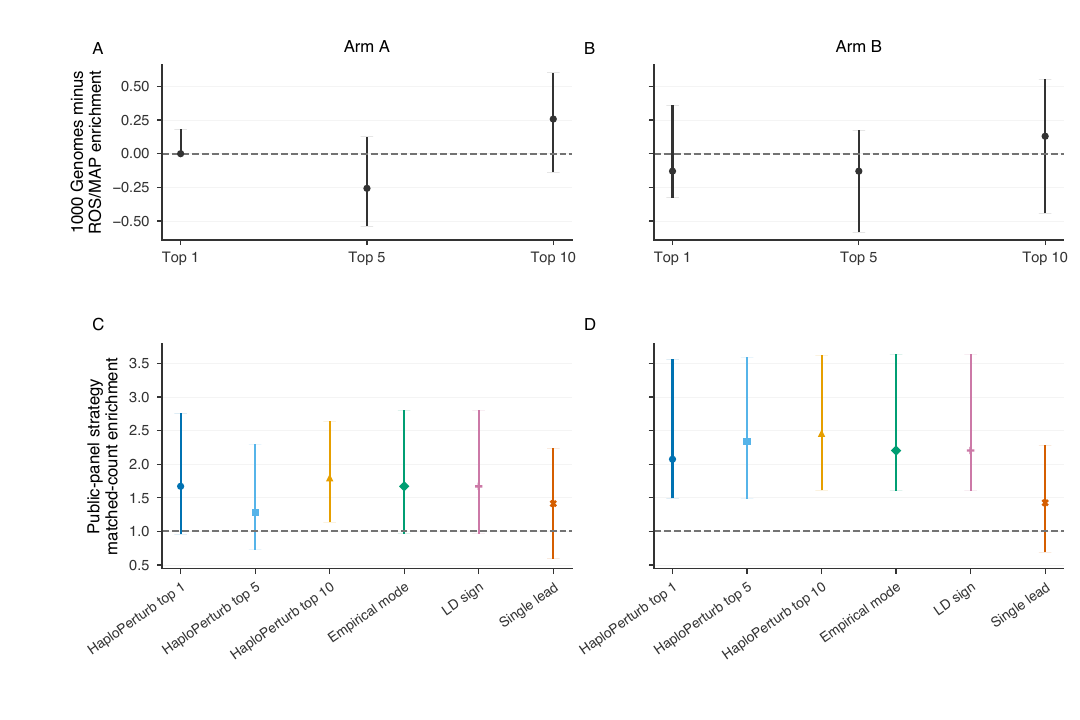}
\caption{\textbf{Public-panel haplotypes preserve microglial eQTL enrichment and perform similarly
under HaploPerturb and simpler construction strategies.} Columns show Arm A ($r^2\geq0.8$) and Arm B
($r^2\geq0.5$). \textbf{(A,B)} Paired 1000 Genomes-minus-ROS/MAP enrichment differences for
HaploPerturb top-one, top-five and top-ten summaries; positive values favor 1000 Genomes. The
panel-specific enrichment estimates underlying these differences are shown in Supplementary
Figure~7. \textbf{(C,D)} Enrichment for public-panel HaploPerturb top one, top five and top ten, the
empirical conditional mode, the LD-sign background and the single-lead strategy. All scores use
complete gene-level exon unions. Error bars are the $2.5$th and $97.5$th percentiles from $10{,}000$
whole-locus bootstrap samples. Dashed lines mark difference zero in panels (A,B) and enrichment one
in panels (C,D). Every public-minus-donor interval crosses zero. In Arm B, all HaploPerturb and simpler
public-panel haplotype intervals lie above one, whereas the single-lead interval crosses one.}
\label{fig:reference-panel-eqtl}
\end{figure}

The common comparison universe remained the same as in Section~\ref{sec:internal-donor-eqtl}: Arm A contained $M=176$ locus--gene pairs and $K=37$ Fujita positives, whereas Arm B contained $M=251$ and $K=44$. In Arm A, the 1000 Genomes top-one, top-five and top-ten rankings recovered 13, 10 and 14 positives, giving enrichments $1.67$ $[0.95,2.76]$, $1.29$ $[0.73,2.30]$ and $1.80$ $[1.14,2.64]$, respectively (Figure~\ref{fig:reference-panel-eqtl}C). The corresponding ROS/MAP enrichments were $1.67$, $1.54$ and $1.54$ (Supplementary Figure~7A). The paired 1000 Genomes-minus-ROS/MAP differences were $0.00$ $[0.00,0.18]$, $-0.26$ $[-0.54,0.13]$ and $0.26$ $[-0.14,0.60]$ (Figure~\ref{fig:reference-panel-eqtl}A). Thus none excludes zero. Although the public-panel top-ten interval excludes one, its paired difference from the single-lead enrichment is $0.39$ $[-0.20,1.23]$, so Arm A does not establish an advantage over the single-lead perturbation.

In Arm B, the 1000 Genomes top-one, top-five and top-ten rankings recovered 16, 18 and 19 positives. Their enrichments were $2.07$ $[1.49,3.56]$, $2.33$ $[1.49,3.59]$ and $2.46$ $[1.61,3.62]$; all three intervals exclude one (Figure~\ref{fig:reference-panel-eqtl}D). The corresponding ROS/MAP enrichments were $2.20$, $2.46$ and $2.33$ (Supplementary Figure~7B), and the paired public-minus-donor differences were $-0.13$ $[-0.33,0.36]$, $-0.13$ $[-0.58,0.18]$ and $0.13$ $[-0.44,0.55]$ (Figure~\ref{fig:reference-panel-eqtl}B). Again, none excludes zero. Relative to the single-lead enrichment of $1.43$, however, the public-panel advantages were $0.65$ $[0.23,1.92]$, $0.91$ $[0.17,2.01]$ and $1.04$ $[0.24,2.08]$, all excluding zero.

We also compared public-panel HaploPerturb top one with the empirical-mode and LD-sign strategies from
Section~\ref{sec:simple-haplotype-baselines}, reusing the existing AlphaGenome predictions for the
identical sequence or its mapped HaploPerturb rank. No new model query was needed. The comparison used the
same locus--gene universe and a single paired whole-locus bootstrap across HaploPerturb top one, empirical
mode, LD sign and single lead. Figure~\ref{fig:reference-panel-eqtl}C--D places those microglial
results beside HaploPerturb top five and top ten, and Supplementary Figure~8 reports the four-strategy
comparison across all three cell types.

In Arm A microglia the three haplotype-aware strategies are identical and each has enrichment $1.67$
$[0.96,2.80]$, compared with $1.41$ $[0.60,2.24]$ for single lead; their paired advantage of $0.26$
$[-0.44,1.42]$ does not exclude zero. In Arm B, HaploPerturb top one recovers 16 positives and has
enrichment $2.07$ $[1.48,3.60]$. The identical empirical-mode and LD-sign rankings recover 17 and
have enrichment $2.20$ $[1.60,3.64]$, whereas single lead recovers 11 and has enrichment $1.43$
$[0.69,2.29]$. Both haplotype-aware comparisons with single lead exclude zero: $0.65$
$[0.23,1.89]$ for HaploPerturb top one and $0.78$ $[0.30,1.95]$ for either simple background. The HaploPerturb
minus empirical-mode difference is $-0.13$ $[-0.35,0.13]$. Thus the factor-model construction does
not outperform the simple haplotype backgrounds in this application; instead, all three show that
specifying a population-informed haplotype rather than a lead-only reference background changes the
microglial ranking.

The additional Arm B positive under the empirical-mode and LD-sign rankings is \emph{KANSL1} at the
diffuse chromosome 17 locus, where HaploPerturb rank two replaces the uncertified HaploPerturb rank-one
alternate-state background. Across the remaining loci the three haplotype-aware scores are
identical. Under HaploPerturb top one, the 16 selected Fujita positives span 12 loci, compared with 11
positives across eight loci under single lead. Haplotype-aware ranking promotes positives at
\emph{PICALM}, \emph{CD2AP}, \emph{CASS4}, and the \emph{MS4A} cluster that the lead-only ranking
does not select (Supplementary Table~2). These are confirmations within the Fujita benchmark, not
new gene discoveries. Their biological coherence is nonetheless useful: experimental studies have
linked the \emph{MS4A} cluster to soluble TREM2 regulation \citep{deming2019ms4a}, shown that
microglial cis-regulatory elements are enriched for AD risk and can be connected to target genes by
CRISPR perturbation \citep{yang2023microglia}, and connected reduced \emph{INPP5D} function in human
microglia to inflammasome activation \citep{chou2023inpp5d}.

A descriptive leave-one-locus-out analysis gave HaploPerturb-top-one enrichment between $1.99$ and $2.46$
after removing any single Arm B locus. The largest downward change, to $1.99$, followed removal of
the \emph{MS4A} locus \texttt{11:60254475:G:A}; removing the unstable chromosome 17 locus increased
enrichment to $2.46$. The aggregate pattern is therefore distributed across loci and is not driven
by the one construction failure. Because this diagnostic recomputes point estimates rather than a
new inferential analysis, it should be read as influence assessment, not as a separate replication.

The cross-panel haplotype differences are documented in Section~\ref{sec:cohort-haplotype-agreement}. Despite those differences, the 1000 Genomes-derived sequences preserve the Arm B microglial enrichment and its advantage over a lead-only edit, without a detectable loss relative to sequences constructed from paired ROS/MAP genomes. For these European-ancestry AD loci, this supports using the public panel for eQTL prioritization when donor-specific whole-genome sequencing is unavailable, subject to the public-list certification caveat in Section~\ref{sec:cohort-haplotype-agreement}.

\section{Discussion}
\label{sec:discussion}

The main practical implication is that the full sequence supplied to a sequence-to-function model should be treated as an analysis choice rather than an invisible preprocessing default. In a lead-only edit, every linked but unedited allele is implicitly fixed to its reference-genome state, thereby specifying a particular joint haplotype that may be rare or absent in the target population. Because a sequence-to-function model reads the entire input window, this hidden choice can change predicted effects and variant rankings even when the prediction model itself is held fixed. The central contribution of our framework is to make the input background explicit and replace an arbitrary reference-based configuration with haplotypes informed by phased population data. A lead-only edit remains a useful comparator, but it should not automatically be regarded as the population-representative input. The reference panel, partner definition, construction method and treatment of all unmodeled bases should therefore be stated explicitly.

The haplotype panel should be matched as closely as possible to the ancestry and sampling population of the target study. Sufficiently large cohort-specific phased genotypes offer the closest population match, whereas a well-matched public panel is a practical alternative when cohort genotypes are unavailable. Partner thresholds should be selected before downstream prediction and examined through sensitivity analyses, because the present comparison does not identify an optimal $r^2$ cutoff. A one-factor HaploPerturb fit is a reasonable starting point after high-LD partner selection, but adequacy should be checked using both variance concentration and residual allelic correlation. Loci with $p>n$, visible residual blocks or diffuse candidate probabilities should be flagged for a richer model or more conservative reporting. If the HaploPerturb top configuration, empirical conditional mode and LD-sign background agree and only one background is needed, that agreement should be reported and a simpler strategy may suffice. The HaploPerturb distribution remains useful when several backgrounds, probability weighting or uncertainty assessment are required. The rank-one background is appropriate when it clearly dominates and candidate coverage is certified; otherwise, several leading backgrounds should be retained. Reports should include the candidate-generation strategy, simulation size and seed, HaploPerturb-probability rescoring method, and coverage check. Downstream strategy comparisons should use the same tested universe, give the underlying positive counts as well as enrichment, and use paired resampling of whole loci.

Several alternatives may be preferable in particular settings. Directly using observed phased haplotypes avoids distributional modeling when the reference panel is large enough for exact configurations to recur, but provides poor coverage as the partner dimension grows. Deterministic factor-space enumeration is efficient and reproducible for a one-factor model with moderate $p$; conditional sampling is more flexible for large loci, multiple factors or candidates separated by several allele changes, provided discovery coverage is assessed. When one factor leaves structured residual dependence, multi-factor or block-specific models are natural extensions. Finally, the equal top-$L$ averages used here could be compared with HaploPerturb-probability-weighted averages or explicit marginalization over haplotype uncertainty. These summaries target different quantities and should be chosen prospectively rather than selected by their downstream benchmark performance.

\paragraph*{Limitation} The analysis covers one disease, 38 loci, predominantly European-ancestry panels and one sequence-to-function model. The partner sets were defined in 1000 Genomes, the molecular benchmark came from ROS/MAP, and the ROS/MAP-derived construction and eQTL truth are therefore not an independent replication. The eQTL analysis evaluates group-level ranking enrichment; it does not discover individual eQTLs, establish causal variants or validate predicted effect directions. Lack of enrichment in astrocytes and excitatory neurons cannot be attributed uniquely to AlphaGenome because biosample correspondence, eQTL power and RNA-summary choice may also contribute. The locus bootstrap captures variation across the sampled GWAS windows but not uncertainty in AlphaGenome, which is deterministic only conditional on a supplied sequence. Reproducibility is further limited by the absence of a permanent identifier for the hosted server-side model. At the construction stage, weakly linked and unavailable variants are omitted, lower-ranked haplotypes are poorly resolved by finite panels, and the largest diffuse locus shows that neither a one-factor approximation nor finite candidate sampling is uniformly reliable.

The present inputs remain partial haplotypes because HaploPerturb alleles are written onto GRCh38 while all bases outside the lead and selected partner set remain at the reference state. Haplotype-resolved assemblies from a human pangenome could instead provide a coherent sequence across the full predictor window, including unmodeled alleles and structural variants \citep{wang2022humanpangenome}. This would still require statistical selection of a population- or individual-appropriate path and consistent mapping of graph coordinates and annotations to the prediction model. Multi-population analyses should additionally separate within-ancestry LD from dependence induced by ancestry-specific allele frequencies; admixed genomes require local as well as global ancestry information, and heterogeneous LD may require partially pooled ancestry-specific loadings. Evaluation across additional diseases, ancestries, prediction models and independent molecular-trait cohorts will be needed to determine when these more complete constructions improve functional prioritization.

\section*{Supplementary material}

The supplement contains estimation and diagnostic details; the conditional sampler, candidate
enumeration, probability evaluation and coverage checks; the proof of one-factor identifiability;
the known-truth finite-panel validation; the paired whole-locus bootstrap; and additional figures
and locus--gene tables.

\section*{Funding and support}

This work was supported by the National Human Genome Research Institute of the National Institutes
of Health under Award Number R01HG012555. The content is solely the responsibility of the author and
does not necessarily represent the official views of the National Institutes of Health. The author
gratefully acknowledges the participants and investigators of the Religious Orders Study, the Rush
Memory and Aging Project, the 1000 Genomes Project, and the contributing studies of the AD Knowledge
Portal.

\section*{Data and code availability}

The 38 Alzheimer's disease lead variants are listed by \citet{wightman2021genome}; the article and
its supplementary information are available at
\url{https://doi.org/10.1038/s41588-021-00921-z}. The 1000 Genomes 30$\times$ high-coverage GRCh38
release is publicly distributed by the International Genome Sample Resource, with collection
metadata at \url{https://www.internationalgenome.org/data-portal/data-collection/1000genomes_30x/}
and the chromosome-specific phased VCFs used here available from
\url{https://ftp.1000genomes.ebi.ac.uk/vol1/ftp/data_collections/1000G_2504_high_coverage/working/20220422_3202_phased_SNV_INDEL_SV/}.

The ROS/MAP WGS VCFs used for the donor-genotype analysis are access-controlled at AD Knowledge
Portal/Synapse accession \texttt{syn11724057}
(\url{https://www.synapse.org/Synapse:syn11724057}). Qualified investigators may request access by
following the \href{https://adknowledgeportal.synapse.org/Data\%20Access}{AD Knowledge Portal
data-access procedures}; individual-level genotypes cannot be redistributed with this article. The
Fujita et al. cell-type-level eQTL summary statistics are available at Synapse accession
\texttt{syn52335732} (\url{https://doi.org/10.7303/syn52335732}), and the underlying DLPFC
single-nucleus RNA-seq data are available under accession \texttt{syn31512863}
(\url{https://www.synapse.org/Synapse:syn31512863}), subject to the applicable data-use terms.

Two GitHub repositories accompany this paper. The reusable HaploPerturb software, user tutorial and
worked examples are available at
\url{https://github.com/jichunxie/haploperturb}. The complete paper reproducibility archive is
available at
\url{https://github.com/jichunxie/haplotype-reproducibility}; it contains versioned analysis
scripts, public-panel manifests, simulation code, seeds, environment records and the sanitized
JSON/CSV artifacts underlying the reported results. Code in both repositories is released under
the MIT License. Protected donor-level genotypes and hypothesis-level protected prediction results
are not included; the reproducibility repository documents the controlled-access stages and the
aggregate public reproduction path.

\paragraph*{AlphaGenome predictor and reproducibility}
AlphaGenome converts each constructed sequence contrast into a predicted transcriptomic effect. We queried its hosted API \citep{avsec2026advancing} with \texttt{alphagenome 0.4.0}, requesting human \texttt{RNA\_SEQ} output for 1,048,576-bp (approximately 1.05-Mb) GRCh38 inputs. All perturbation strategies were evaluated using the same AlphaGenome RNA-seq tracks. We recorded the AlphaGenome client version, analysis scripts, requested output type, and track metadata. However, because the hosted API did not report a permanent version identifier for the server-side model, an exact future rerun may depend on whether that model has since been updated.

\bibliographystyle{plainnat}
\bibliography{references}

\end{document}